\documentclass[acmsmall,screen]{acmart}

\usepackage{CJKutf8}
\usepackage[utf8]{inputenc}
\usepackage[T1]{fontenc}
\usepackage{textcomp}
\usepackage{xcolor,colortbl}
\usepackage[most]{tcolorbox}
\usepackage{multirow}
\usepackage{multicol}
\usepackage{threeparttable}
\usepackage{soul}
\usepackage{hyperref} 
\usepackage{longtable}
\usepackage{subfig}
\usepackage{overpic}
\usepackage{array}
\usepackage{float}
\usepackage{tabularx}
\usepackage[english]{babel}
\usepackage{booktabs}
\usepackage{hyphenat}
\newcommand{\modtext}[1]{\textcolor{black}{#1}}
\newcommand{\revtext}[1]{\textcolor{black}{#1}}

\AtBeginDocument{
  \providecommand\BibTeX{{
    \normalfont B\kern-0.5em{\scshape i\kern-0.25em b}\kern-0.8em\TeX}}}

\newif\ifdraft
\drafttrue

\renewcommand\footnotetextcopyrightpermission[1]{} 
\begin{document}
\setcopyright{acmcopyright}
\copyrightyear{2025}
\acmYear{2025}
\acmDOI{XX.XXXX/XXXXXXX.XXXXXXX}

\acmJournal{JACM}
\acmVolume{1}
\acmNumber{1}
\acmArticle{1}
\acmMonth{1}

\title[Value Sensitive Design for Fair Online Recruitment]{Value Sensitive Design for Fair Online Recruitment: A Conceptual Framework Informed by Job Seekers' Fairness Concerns}

\author{Changyang He}
\authornote{These authors contributed equally to this work.}
\authornote{This work was primarily conducted during the author's stay at the Max Planck Institute for Security and Privacy (MPI-SP).}
\affiliation{%
  \institution{Harbin Institute of Technology, Shenzhen}
  \city{Shenzhen}
  \country{China}
}
\affiliation{%
  \institution{Max Planck Institute for Security and Privacy (MPI-SP)}
  \city{Bochum}
  \country{Germany}
}
\email{hechangy@hit.edu.cn}

\author{Yue Deng}
\authornotemark[1]
\affiliation{%
  \institution{Hong Kong University of Science and Technology}
  \city{Hong Kong}
  \country{China}
}
\affiliation{%
  \institution{Max Planck Institute for Security and Privacy (MPI-SP)}
  \city{Bochum}
  \country{Germany}
}
\email{ydengbi@cse.ust.hk}

\author{Alessandro Fabris}
\affiliation{%
  \institution{University of Trieste}
  \city{Trieste}
  \country{Italy}
}
\affiliation{%
  \institution{Max Planck Institute for Security and Privacy (MPI-SP)}
  \city{Bochum}
  \country{Germany}
}
\email{alessandro.fabris@units.it}

\author{Bo Li}
\affiliation{%
  \institution{Hong Kong University of Science and Technology}
  \city{Hong Kong}
  \country{China}
}
\email{bli@cse.ust.hk}

\author{Asia J. Biega}
\affiliation{%
  \institution{Max Planck Institute for Security and Privacy (MPI-SP)}
  \city{Bochum}
  \country{Germany}
}
\email{asia.biega@mpi-sp.org}

\renewcommand{\shortauthors}{He et al.}

\newcommand{\gr}{\leavevmode\color{lightgray}}
\newcommand{\cgr}{\cellcolor{green!25}}
\newcommand{\crd}{\cellcolor{pink!25}}

\begin{abstract}

The susceptibility to biases and discrimination is a pressing issue in today's labor markets. While digital recruitment systems play an increasingly significant role in human resource management, a systematic understanding of human-centered design principles for fair online hiring remains lacking, particularly considering the gap between idealized conceptualizations of fairness in research and actual fairness concerns expressed by job seekers. To address this gap, this work explores the potential of developing a fair recruitment framework based on job seekers' fairness concerns shared in r/jobs, one of the largest online job communities. Through a grounded theory approach, we uncover four overarching themes of job seekers' fairness concerns: \revtext{personal attribute discrimination beyond legally protected attributes}, interaction biases, improper interpretations of qualifications, and power imbalance. Drawing on value sensitive design, we derive design implications for fair algorithms and interfaces in recruitment systems, integrating them into a conceptual framework that spans different hiring stages.

\end{abstract}



\maketitle

\section{INTRODUCTION}

\revtext{Given the growing scale and complexity of modern recruitment}, online recruitment plays an increasingly important role in human resource management. Online recruitment systems connect recruiters and job seekers at scale, and afford efficient sourcing, screening, assessment, selection, and hiring of candidates~\cite{lashkari2023finding,dillahunt2021examining,fabris2023fairness}. Such systems often \modtext{incorporate} automated tools to streamline the hiring process, such as filtering and ranking interfaces to assist hiring decisions~\cite{dillahunt2021examining}, allowing scalable and efficient recruiter-candidate matching.





\revtext{Despite the potential of online recruitment, ensuring fair hiring remains a critical challenge in such systems. Extensive evidence indicates that online recruitment frequently perpetuates and can even amplify discrimination, bias, and imbalances present in offline hiring}~\cite{nagaraj2023discrimination, salinas2023unequal,suhr2021does,hannak2017bias,lashkari2023finding}. In particular, online recruitment introduces new fairness dynamics through the combined influence of \textit{algorithm} and \textit{interface} design. On the algorithmic side, for instance, candidate ranking algorithms, \revtext{which are predominantly data-driven}, may produce outcomes that favor historically advantaged groups due to representation bias in the training data~\cite{fabris2023fairness}. Similarly, job recommendation models might exhibit biases against various demographic identities in recommended job types~\cite{salinas2023unequal}. Regarding interface design, a typical case is that social feedback interfaces might exacerbate biases against protected gender and race groups in the freelancing platform, as they may receive worse reviews given equal qualifications, hindering their future work opportunities~\cite{leung2020race}. Moreover, the interactions between the hiring algorithms and interfaces can further compound unfair treatment. For example, the selective use of demographic imagery in job advertisements, combined with biased ad delivery algorithms, may jointly marginalize underrepresented job seekers~\cite{nagaraj2023discrimination}.



In response to these fairness challenges, prior research has proposed both algorithmic and interface-level interventions. To address biases in hiring algorithms, researchers in responsible computing have developed diverse fairness measures for algorithm optimization (e.g., outcome fairness~\cite{hardt2016equality} and process fairness~\cite{grgic2016case}), as well as pre-, in-, and post-processing approaches for bias mitigation~\cite{fabris2023fairness,langenkamp2020hiring}. Regarding interface design for fair hiring, scholars in Human-Computer Interaction (HCI) and Computer-Supported Cooperative Work (CSCW) have investigated design spaces to reduce human bias and assist fair decision-making, such as individual-level information provision~\cite{leung2020race} and human-AI collaboration in debiasing~\cite{echterhoff2022ai,lashkari2023finding}.

While the research progress in fair online hiring is promising, \textit{whether these efforts adequately align with job seekers' actual fairness concerns} remains an open question. In fact, recent research has highlighted the gap between idealized fair hiring models and the manifestations and perceptions of fairness in real life~\cite{sarkar2024s,zhang2022examining,lavanchy2023applicants,cave2019hopes}. Computational approaches to bias mitigation often risk oversimplification, such as neglecting the intersectionality of social identities~\cite{sarkar2024s,fabris2023fairness}. Additionally, different notions of fairness may reflect fundamentally divergent worldviews~\cite{friedler2021possibility}, and specific operationalizations of fairness may not generalize across settings, potentially perpetuating existing inequalities~\cite {sarkar2024s}. Moreover, fairness is not a binary concept~\cite{anderson2011perceivedLong}. Job applicants may perceive algorithmic recruitment as unfair, even when outcomes are in their favor~\cite{lavanchy2023applicants}. In some cases, this perception stems from concerns about systems failing to recognize candidates' uniqueness~\cite{lavanchy2023applicants} or from a lack of transparency in decision-making~\cite{armstrong2023navigating}. Given these concerns, \textbf{adopting a human-centered approach to fair hiring} is crucial for translating real-world fairness challenges into implications for fair online hiring designs. It not only necessitates the refinement of \textit{algorithm design} to better correspond to job seekers' fairness needs, but also highlights the significance of \textit{interface design} to enhance fairness perceptions beyond theoretical and computational fairness notions.

This paper takes a first step toward building a human-centered framework for fair online hiring, grounded in job seekers' concerns. Specifically, we investigate the following questions:

\begin{itemize}
  
  \item \textbf{RQ1}: Which fairness concerns are raised by job seekers in online job communities?

  \item \textbf{RQ2}: How might job seekers' fairness concerns translate into design insights for online hiring systems, in particular recruitment algorithms and interfaces?

\end{itemize}

For RQ1, we applied text classification to identify posts related to job seekers' fairness concerns in r/jobs\footnote{https://www.reddit.com/r/jobs/}, one of the largest online job communities~\cite{garg2021using}, and used a grounded theory approach to inductively analyze fairness concerns in these posts. Our findings suggest four significant themes of job seekers' fairness concerns, covering \textit{\revtext{personal attribute discrimination beyond legally protected attributes}}, \textit{interaction bias}, \textit{improper interpretations of qualifications}, and \textit{power imbalance}. These fairness concerns offer a comprehensive overview of hiring challenges in the real world, shedding light on fairness perspectives that job seekers highlight but are potentially overlooked in the existing algorithmic fairness literature.

For RQ2, we examined how practitioners can translate the broad range of fairness considerations into concrete responsible hiring system development. Building upon the multifaceted fairness concerns identified in RQ1, we followed value sensitive design \modtext{(VSD)}~\cite{friedman1996value} to develop design implications for algorithms and interfaces across different hiring stages and contexts. For example, we suggest enhancing \textit{two-sided fairness} to address power asymmetry in job recommendations, and adopting \textit{visual feature anonymization} to mitigate biases resulting from visual proxies. These directions offer a conceptual foundation for future work to implement, evaluate, and operationalize fair hiring interventions.


In summary, this work makes the following contributions to the HCI, CSCW, and responsible computing community: (1) we construct a comprehensive taxonomy of job seekers' fairness concerns in hiring, which forms the foundation of user-centered design for online hiring systems; (2) we outline a conceptual framework for designing fair recruitment \textit{interfaces}, which situates algorithmic fairness within broader real-world fairness challenges; (3) we propose novel directions for designing fair recruitment \textit{algorithms}, which aim to align theoretical and computational fairness with users' actual fairness perceptions.

\section{RELATED WORK}

\subsection{Understanding job seekers' fairness concerns in online hiring}

With the turbulent market and intense competition in today's economy, effective human resource management that aims to maximize candidate-job \modtext{matching} plays an irreplaceable role in business success~\cite{alder2006achieving,pfeffer1998human}. Unfortunately, although there have been substantial legal~\cite{USLaw,EuroLaw}, social~\cite{skrentny2006policy,sanchez2020does}, and technical~\cite{fabris2023fairness,raghavan2020mitigating} efforts for fair and non-discriminatory hiring, job seekers still hold broad fairness concerns when interacting with the recruitment ecosystem~\cite{lashkari2023finding,zhang2022examining}. Gaining insight into the fairness concerns of individual job seekers could illuminate strategies for integrating fairness-focused design into the hiring process.

With the advancement of hiring algorithms, HCI and responsible computing researchers have been paying increasing attention to job seekers' fairness concerns in the context of automated hiring. For example, Zhang et al. found that the public generally has a negative attitude toward fairness in hiring algorithms~\cite{zhang2022examining}. Lavanchy et al. also noted that job applicants perceive algorithmic recruitment processes as less fair, regardless of whether the outcome is favorable to the applicant~\cite{lavanchy2023applicants}. Such perceptions may vary according to the identity of job seekers (e.g., more negative attitudes from women compared to men) and the type of algorithms (e.g., more negative attitudes towards video screening algorithms than resume screening algorithms)~\cite{zhang2022examining}. Some reasons for users' fairness concerns about algorithmic hiring include uncertainty \modtext{about} the current technology capability~\cite{lashkari2023finding}, worries about limited transparency in decision-making~\cite{armstrong2023navigating}, and distrust of algorithms' ability to recognize their uniqueness~\cite{lavanchy2023applicants}. Echoing the gap between perceived job discrimination and actual job discrimination~\cite{anderson2011perceivedLong}, these studies warn that despite the progress in outcome fairness of hiring algorithms~\cite{fabris2023fairness,veldanda2023emily}, it is important to critically reflect on more diverse dimensions of system and algorithm design to promote users' perceived fairness in automated hiring systems.

Moving beyond identifying existing limitations to improve recruitment systems and algorithms, this work takes a holistic view to unpack job seekers' fairness perceptions: what fairness concerns do job seekers disclose, and how do these concerns shed light on the design of future hiring systems? To achieve this goal, we situate this study in a natural setting, specifically an online job community, to identify fairness concerns based on job seekers' discourse. Furthermore, we adopted VSD~\cite{friedman1996value} to systematically translate job seekers' broad fairness concerns into actionable design implications. VSD provides a structured approach that involves \textit{empirical investigation} to unpack job seekers' fairness concerns, \textit{conceptual investigation} to surface normative fairness-related values, and \textit{technical investigation} to capture value misalignments in current technologies~\cite{friedman1996value,zhu2018value}. \revtext{In CSCW and HCI communities, VSD has been widely adopted to integrate human values into algorithmic system design~\cite{zhu2018value,sadek2024guidelines,iqbal2021search,sadek2024designing,boyd2022designing,chen2022practitioners,showkat2023right} such as recommender systems~\cite{chen2022practitioners} and socialization algorithms~\cite{zhu2018value}. Recent work has also applied VSD to promote responsible AI in algorithmic systems~\cite{sadek2024guidelines,iqbal2021search,boyd2022designing}}. Therefore, our work complements existing literature by systematically embedding fairness values into fair hiring design, addressing concerns that might otherwise be overlooked in empirical studies due to the opacity of algorithmic decision-making and the limited public understanding of (automated) hiring systems.


\subsection{Designing algorithms for fair online hiring}\label{Algo}

Hiring algorithms have permeated different stages of human resource management. They span stages from \textit{sourcing} to find potential candidates~\cite{nandy2022achieving}, \textit{screening} to narrow down pools of applicants~\cite{elbassuoni2020fairness,lippens2024computer}, \textit{selection} to optimize job offers~\cite{suen2023comparing}, and \textit{evaluation} after hiring~\cite{yaneva2018employee}. As such, biases of hiring algorithms manifest in diverse patterns across different stages. For instance, Rao and Korolova have noted that during job advertising, there is disproportionate representation or exclusion of certain demographics in job ad images, and ad delivery algorithms could further amplify the skews~\cite{nagaraj2023discrimination}. Ranking algorithms employed by online recruitment platforms for candidate selection are also prone to a variety of undesirable biases (e.g., pro-male outcomes)~\cite{suhr2021does,hannak2017bias}.

Facing such problems, HCI and responsible computing researchers have devoted \modtext{considerable effort} to improving fairness in algorithmic hiring. First, the research community of fair hiring has defined different dimensions of fairness measures for algorithm optimization, such as outcome fairness~\cite{hardt2016equality} (equal outcome in candidate predictions), accuracy fairness~\cite{beutel2019fairness} (equal accuracy-related metrics between groups), and representation fairness~\cite{abbasi2019fairness} (stereotyping and biases in representations). Notably, most existing work takes group fairness as the primary objective, while individual fairness remains less frequently adopted as a fairness criterion~\cite{fabris2023fairness}. Besides, recent research has investigated process fairness by examining the predictability of sensitive attributes from non-sensitive ones~\cite{grgic2016case}. To address these fairness criteria, researchers have developed various pre-, in-, and post-processing bias mitigation algorithms. For instance, rule-based removal~\cite{parasurama2021degendering} and substitution~\cite{rus2022closing} as pre-processing techniques attempt to replace words that explicitly refer to sensitive attributes in order to improve outcome fairness. With similar proxy-reduction goals, in-processing Adversarial Inference approaches aim to reduce sensitive information in the latent representation by incorporating an additional adversarial loss~\cite{pena2020bias}. DetGreedy is a typical example of post-processing, which greedily selects the most relevant candidate in the ranking while maintaining maximum and minimum representation constraints for each group~\cite{geyik2019fairness}.

While extensive research efforts have been made toward developing fair hiring algorithms, discrepancies still persist between algorithm development and their real-world implementation. Limited datasets, characterized by issues such as inadequate diversity and the absence of sensitive attributes~\cite{fabris2023fairness}, pose a significant challenge to the generalizability of fair hiring algorithms. \modtext{Additionally}, fair hiring algorithms often oversimplify the recruitment model, overlooking the interplay between different sensitive variables and the cascading effect across multiple hiring stages~\cite{langenkamp2020hiring}. More importantly, due to a lack of model transparency, users often hold misperceptions of how algorithms work and distrust algorithm-based decision-making~\cite{cave2019hopes}.

Following this line of work, our study aims to take a human-centered perspective to identify the potential gap between the advancement of fair hiring algorithms and users' fairness considerations in the hiring pipeline. Drawing insights from users' fairness perceptions, we reflect on what critical aspects future fair hiring algorithms should particularly examine and address.

\subsection{Designing interfaces for fair online hiring}\label{Interface}

As online recruitment is often powered by algorithms to afford automation and scalability, a vast literature exists on how algorithms can inherit or amplify hiring discrimination~\cite{langenkamp2020hiring}. The connection between hiring fairness and the interface design of recruitment systems, by contrast, remains relatively underexplored~\cite{leung2020race}. In fact, as a bridge between humans and algorithms in nearly every decision-making stage for both recruiters and job seekers~\cite{leung2020race,suhr2021does}, the interface design of recruitment systems can play a significant role in influencing (un)fair hiring decisions~\cite{leung2020race}.

One strand of work in HCI and CSCW has explored the relationship between the specific interface design of hiring systems and fairness and biases~\cite{leung2020race,hannak2017bias,chua2020you}. For example, Leung et al. found that the provision of candidate information at the individual level (performance of candidates in previous tasks) can significantly alleviate discrimination, while the provision of information at the subgroup level (how the candidate's subgroup did in previous tasks) can increase discrimination~\cite{leung2020race}. In the setting of online freelancing, Hann\'{a}k et al. found that biases against protected gender and race groups manifest through the social feedback interface, and suggested more selectively revealing review information as a mitigation approach~\cite{hannak2017bias}. Affording accessible recruitment interfaces is another important component for fair hiring in a broad sense~\cite{wang2024there,migovich2024stress,ara2024collaborative}. For instance, Wang et al. revealed how short video and live-streaming interfaces provide a user-friendly channel to aging job seekers~\cite{wang2024there}.

Another thread of HCI and CSCW research focuses on developing interfaces to facilitate fair hiring decisions for recruiters~\cite{echterhoff2022ai,lashkari2023finding,yang2024fair}. For instance, Lashkari and Cheng suggested that recruitment decision-making tools should support team collaboration that allows multiple roles to contribute to the decision process, and should also facilitate (e.g., summarizing candidate profiles) rather than perform decision-making~\cite{lashkari2023finding}. Yang et al. developed and evaluated an AI system adopting fairness-aware machine learning to provide guidance on unbiased decision-making, illustrating its potential for prompting decision-makers to reflect on their biases and reassess fairness-related views~\cite{yang2024fair}. 

This work contributes to the research area by proposing a fair recruitment design framework based on job seekers' fairness concerns. By integrating users' fairness concerns across diverse dimensions, this framework captures nuanced design considerations across different hiring stages.

\section{METHOD}

In this section, we describe our method for creating a fair online recruitment framework based on job seekers' fairness concerns. We first introduce the data source and data collection method in Section \ref{data}. Then, we illustrate how we identified users' fairness concerns through qualitative analysis in Section \ref{RQ1-method}, and developed design insights for online hiring systems to establish the framework in Section \ref{RQ2-method}. We discuss the ethical considerations in Section \ref{ethical_considerations}.

\subsection{Data source and collection}\label{data}

\revtext{
To uncover job seekers' general concerns about fairness in the recruitment process, we chose to examine online job-seeking communities, which allow us to identify job seekers' fairness concerns from a large and diverse group in a natural setting. 
}

\subsubsection{Data source: r/jobs}

To locate online communities where job seekers gather to discuss fairness concerns in job hunting, we first broadly searched popular social media platforms, including Twitter, Facebook, and Reddit, using relevant search keywords (e.g., "hiring bias," "fair hiring," and "recruitment discrimination") in March 2024. We noticed that most posts on Twitter and Facebook public pages \modtext{shared} news in technology and law development on recruitment fairness, yet few shared personal experiences and concerns related to fairness during job seeking. Hence, we decided to focus \modtext{our research} specifically on Reddit, where extensive conversations revolved around fairness concerns as perceived by job seekers and based on their experiences~\cite{garg2021using}. Although there were various subreddits discussing topics related to recruitment fairness, some concentrated particularly on specific stages in job hunting (e.g., ``r/resumes'' and ``r/interviews'' focusing on resume construction and job interviews separately). Therefore, we finally located the subreddit "r/jobs", an online community that allows general job seekers to share opinions and experiences across different job-hunting stages.

The subreddit r/jobs, created in March 2008, is described as ``the number one community for advice relating to your career''\footnote{https://www.reddit.com/r/jobs/}. As of August 2024, r/jobs had 1.6 million members. The community forbids job posts or self-promotion of any kind. Instead, it encourages community members to share advice related to its core mission - supporting topics such as "how to get a job" or "how to quit a job." The community offers a broad range of "flairs" to tag each post, covering topics such as "Job Searching," "Applications," "Resumes/CVs," "Interviews," "Job Offers," "Leaving a Job," and more. Without any specific limitations on job-seeking stages or topics, r/jobs serves as a community where job seekers can exchange a wide range of experiences and opinions, including those relevant to hiring fairness.

\subsubsection{Data collection and preprocessing}

\begin{figure}
	\centering
		\includegraphics[scale=.35]{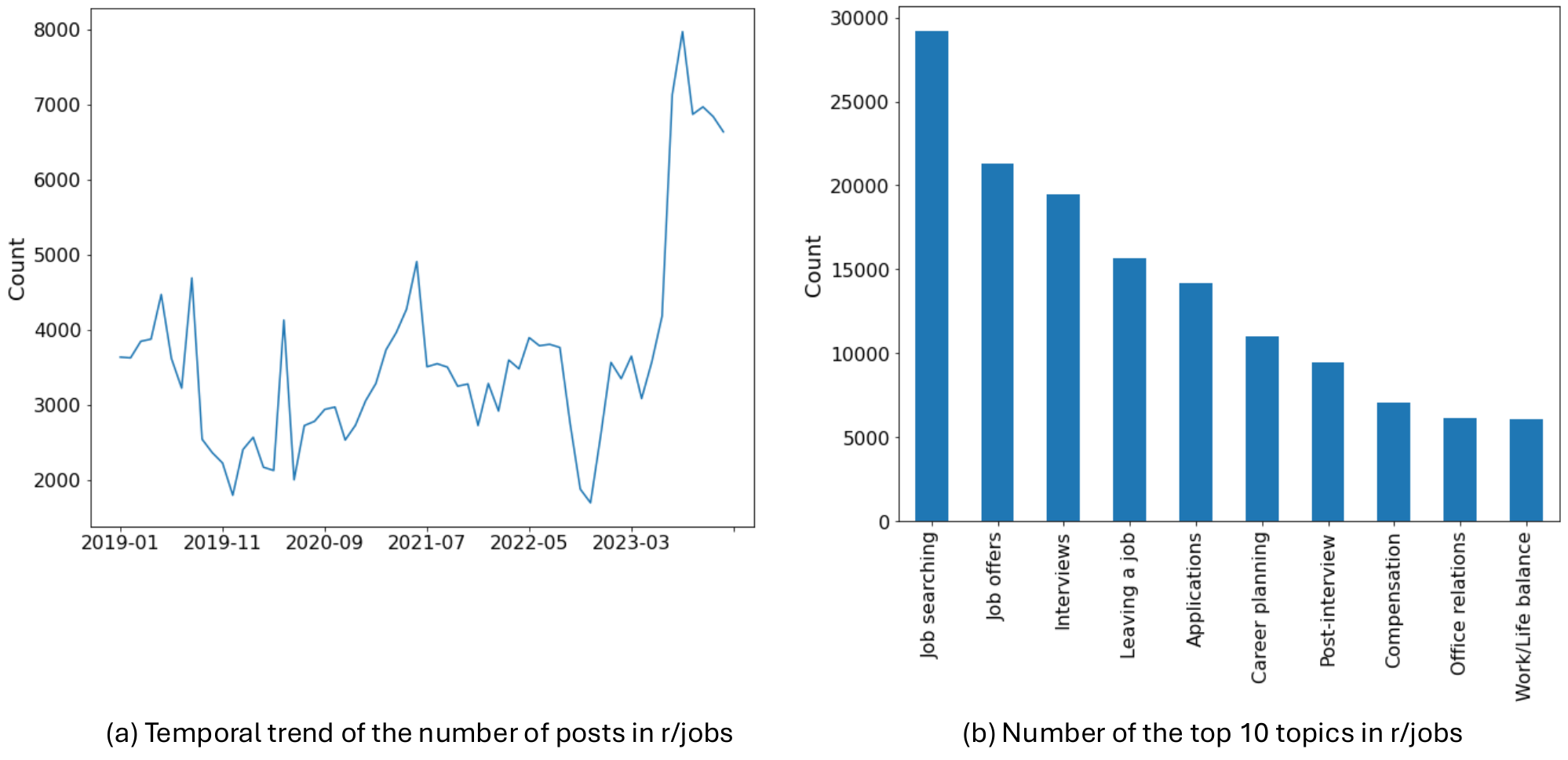}
	\caption{Descriptive statistics on (a) temporal trend and (b) topic distribution of posts in r/jobs. The count in the y-axis indicates the number of posts in the corresponding month(s) for (a) and the number of posts with relevant tags for (b).}
        \Description{Descriptive statistics on (a) temporal trend and (b) topic distribution of posts in r/jobs. The count in the y-axis indicates the number of posts in the corresponding month(s) for Figure (a) and the number of posts with relevant tags for Figure (b). Figure (a) shows that there is an increasing trend of posts in r/jobs after 2023. Figure (b) shows that the top 5 most prevalent topics include job searching, job offers, interviews, leaving a job, and applications.}
	\label{FIG: post_statistics}
\end{figure}

\revtext{Due to changes in Reddit's API access policies since 2023, which limit the feasibility of large-scale academic data collection~\cite{RedditAPI}, we relied on publicly available Reddit archive dump files curated by the Pushshift community~\cite{RedditData,sum2025s,chandrasekharan2017you}. We acknowledge that this data source reflects broader tensions and ongoing debates surrounding platform governance and data access for research~\cite{fiesler2024remember}, and obtaining consent for large-scale community analysis is challenging in this setting. Consequently, we used the data exclusively for this non-commercial academic research, and made careful ethical considerations before data analysis, as detailed in Section \ref{ethical_considerations}.} We first extracted all the posts (N=388,108) from r/jobs between 2019 and 2023. Then, we removed all posts with a 'deleted' or 'removed' status, resulting in 216,441 post submissions. The metadata of posts included \textit{id, post title, post description, created timestamp, and hashtag}, as well as \textit{number of comments, views, and votes (ups/downs)}. The temporal trend and topic distributions of posts in r/jobs are shown in Figure \ref{FIG: post_statistics}.

\subsection{RQ1: Identifying users' fairness concerns}\label{RQ1-method}

\subsubsection{Conceptualizing fairness} Considering the complexity of fairness, we grounded our evaluation of \textit{whether a post contained users' fairness concerns} in the theoretical framework proposed by Colquitt and Rodell~\cite{colquitt2015measuring}. This framework assumes a broad definition of ``fairness'', which aligns with our research goal to surface a comprehensive taxonomy of users' fairness concerns. According to this framework~\cite{colquitt2015measuring}, a post would be included for analysis if it was related to any of the following four facets of the job-seeking process: (1) procedural fairness, i.e., the fairness perceived in the procedures used to make decisions; (2) distributive fairness, i.e., perceived fairness of outcomes; (3) interpersonal fairness, i.e., the perception of fairness in interpersonal treatment and interactions; (4) informational fairness, i.e., the transparency and adequacy of information during decision-making processes. The data inclusion criteria enabled us to identify posts that reflected job seekers' various fairness concerns, laying the foundation for further analysis.

\subsubsection{Classifying Fairness-related Posts}

Through exploratory analysis to \modtext{familiarize ourselves with the dataset}, we observed that the r/jobs community covered miscellaneous topics about seeking or quitting a job, thus a random sample might contain a low proportion of posts related to fairness concerns in job-seeking. Therefore, we leveraged text classification to establish a valid dataset for qualitative analysis.

\paragraph{Sampling} Considering the relatively small fraction of fairness-related posts in the dataset, we took a strategic sampling approach to construct the training sample, combining 500 random posts from \textit{the whole dataset} and 500 posts randomly selected from \textit{a keyword-filtered subset}. In particular, we included a post in the \modtext{keyword-filtered} subset if it contained any of the following keywords about fairness: \textit{bias(ed), discriminat(e/ion), prejudice(d), stereotype, (un)fair(ness),  (im)partial(ity), (in/un)equal(lity)}. \modtext{These keywords have been commonly observed as empirical manifestations of fairness in online communities~\cite{pessianzadeh2025exploring} to express unequal and discriminatory treatment, aligning with the general definition of fairness as ``impartial and just treatment or behaviour without favouritism or discrimination'' in the Oxford English Dictionary. We included different morphological forms of these words to capture diverse relevant expressions. However, relying only on keywords to locate fairness-related posts has inherent limitations: it mainly surfaces explicit articulations of fairness (e.g., ``employers often subconsciously discriminate based on names'') while potentially missing implicit expressions (e.g., ``the suit led to the rejection of a more qualified candidate''), and posts containing these terms may occasionally address unrelated issues in our task (e.g., bias in office relationships). To reduce such risks, the keyword-filtered subset was not treated as the ground truth, but only as a means to increase the proportion of relevant samples. Therefore, we supplemented it with 500 randomly sampled posts from the full dataset to compensate for missing themes and balance potential keyword bias, resulting in a total of 1000 posts for annotation.}


\paragraph{Annotation}

We manually annotated the sample to assign a binary label to each post, i.e., whether it was related to users' fairness concerns in job-seeking or not. After conceptualizing fairness, we established the criteria for annotation: a post was labeled as positive if (1) it touched on any of the facets of \textit{procedural, distributive, interpersonal} and \textit{informational} fairness~\cite{colquitt2015measuring}; (2) it focused on fairness in the hiring process, for example, posts criticizing workplace discrimination after hiring were excluded. Two researchers first individually coded the initial 100 samples. Cohen's Kappa (a statistical measure to quantify inter-rater reliability)~\cite{mchugh2012interrater} reached 0.93, suggesting substantial agreement between the two coders. The two coders engaged in multiple rounds of discussions to reconcile discrepancies. Finally, the two researchers individually annotated an additional set of 450 posts each, resulting in a total of 1000 labeled samples.

\paragraph{Model training}
We fine-tuned RoBERTa, a variant of BERT, as the base model for our specific task, considering its effective contextual embedding for text classification~\cite{liu2019roberta}. We divided the 1000 labeled samples into training, validation, and testing sets using an 8:1:1 ratio, and concatenated the post title and description as the text input. Given the limited sample size for fine-tuning, we adopted a dropout rate of 0.2, and employed early stopping to prevent overfitting. The text classification achieved good performance with an F1-score = 82.0\% on the test set\footnote{Through error analysis of false positive posts, we identified several categories of misclassification, including (1) complaints of job seeking not related to fairness perceptions, (2) advice-seeking of job seeking not related to fairness concerns, and (3) fairness concerns in the workplace instead of during job seeking. We excluded the misclassified posts in further analysis of fairness concerns.}. We applied the classifier to assign the label of \textit{whether related to job seekers' fairness concerns in hiring} for all 216,441 r/jobs posts in our dataset. The text classification yielded 28,567 posts reflecting job seekers' concerns about fairness in hiring, accounting for 13.2\% of the entire dataset.

\subsubsection{Analyzing fairness concerns}

The text classification prepared a subset for qualitative analysis of users' fairness concerns. To inductively capture the fairness concerns raised by job seekers, we employed a grounded theory approach~\cite{corbin2015basics} to analyze the data. This allowed the codes to naturally emerge during the analysis process. First, two authors employed open coding~\cite{corbin2015basics} to independently analyze 200 samples, generating initial codes. \modtext{During the coding, they carefully situated users' fairness concerns in the specific hiring context disclosed in the post (e.g., hiring stages, tools, and interaction patterns). For example, they noted how \textit{appearance bias} occurs in diverse stages including CV screening and interviews, and intersects with ageism and sexism.} The two coders subsequently engaged in multiple rounds of meetings, comparisons, and discussions to establish a consensus on the initial codebook. During the coding process, coders excluded misclassified posts to ensure the validity and purity of the codebook. Examples of codes included ``\textit{bias in the communication method}'' and ``\textit{prejudice related to working history}''. Subsequently, the two authors iteratively revisited the data, applied the established codebook to analyze additional posts, and documented any newly emerging codes. The coding phase concluded when no new codes surfaced. Following this iterative coding process, the two authors coded 870 posts in total. The two coders used axial coding~\cite{corbin2015basics} to categorize the codes and merge similar ones into overarching themes, such as merging ``\textit{bias in the communication method}'' and ``\textit{bias in the self-presentation approach}'' into the theme ``\textit{interaction bias}''. Finally, the two authors analyzed 100 additional posts based on the final codebook and no new codes emerged, validating that the coding reached saturation.

\subsection{RQ2: Value sensitive design for online hiring systems}\label{RQ2-method}

Based on the taxonomy of job seekers' fairness concerns, we developed design insights for online hiring systems. \modtext{This process followed principles of VSD~\cite{friedman1996value}, which emphasize integrating human values (fairness in this work) into technology design \revtext{and has been widely adopted to enhance responsible algorithmic systems~\cite{sadek2024guidelines,iqbal2021search,boyd2022designing}}. VSD provides a systematic process covering empirical, conceptual, and technical investigations. It builds a structured bridge, systematically linking fairness concerns to social and institutional values and identifying value misalignments in current recruitment systems for design implications. This framing positions our work not merely as an analysis of online discourse but as a principled framework for embedding fairness values into hiring algorithms and interfaces.}


%

\modtext{
Two authors with expertise in recruitment fairness, but complementary backgrounds (one in HCI/CSCW and one in AI), applied VSD to develop fair online hiring systems. Meanwhile, two other authors with HCI and responsible AI backgrounds contributed to the validation and refinement of these systems through iterative discussions. Specifically, VSD involves:}
\begin{enumerate}
    \item \modtext{\textit{Empirical investigation} (what people value): The authors revisited the findings of RQ1 to deepen their understanding of job seekers' experiences and concerns about recruitment fairness. This investigation examined not only specific fairness issues but also shared patterns (e.g., most concerns under interaction bias reflected the worries about ``decisions based on subjective impressions rather than qualifications'').}
    
    \item \modtext{\textit{Conceptual investigation} (what values are at stake, and why): The authors analyzed the normative dimensions of identified fairness concerns, mapping them to social, cultural, and institutional foundations. For instance, from the empirical finding that ``marginalized groups face barriers in system inputs'', the conceptual investigation identified ``inclusion and accessibility'' as fairness-related values.}
    
    \item \modtext{\textit{Technical investigation} (how design can embody or support these values): Drawing on literature in algorithm and interface design for fair hiring, the authors derived design insights by (1) analyzing recruitment systems to identify value misalignments, (2) proposing design interventions to address these gaps, and (3) reflecting on implementation challenges and potential remedies. For example, given the multidimensional discrimination that spans all hiring stages, the authors identified "limited dataset dimensions and stages in fair hiring" as a value misalignment in existing technologies. They proposed "data donation campaigns" as a corresponding implication, while also highlighting the importance of "consent and privacy protection" in implementation.}
    
\end{enumerate}
Finally, the design insights were contextualized within the hiring ecosystem by linking implications to relevant stages and components of the pipeline, thereby achieving contextual integration~\cite{friedman1996value}. This process informed the framework for designing fair hiring algorithms and interfaces.





\subsection{Ethical considerations}\label{ethical_considerations}

\revtext{
Even though our study, as a secondary analysis of publicly available data, is eligible for self-exemption under our Institutional Review Board (IRB) guidelines},\footnote{\revtext{Although IRB approval was not required, we sought informal guidance and feedback from the ethics review committee.}}\revtext{~we recognized that IRB guidelines may not always suffice for community safety, and the publicness of Reddit data might obscure complex ethical issues~\cite{proferes2021studying,fiesler2024remember}. Contextual ethical considerations are necessary, especially since our research involves potentially sensitive data (e.g., disclosure of discriminatory experiences related to specific group identities), and obtaining consent for large-scale analysis of a previously curated dataset is challenging. Therefore, we carefully reflected on the ethical considerations in Reddit research by Fiesler et al.~\cite{fiesler2024remember}, and implemented various protective measures to safeguard privacy and minimize the risks to the studied Reddit community. These considerations are grounded in the three fundamental research ethics principles of \textit{respect for persons}, \textit{beneficence}, and \textit{justice}, with particular attention to the tension between benefits and risks in the study context.}
\begin{enumerate}  

    \item \revtext{While the community data was used mainly for a beneficial purpose, i.e., promoting fair online recruitment, its analysis and presentation may pose privacy risks, especially the possibility of re-identification.} Therefore, we \revtext{removed} all real usernames or user IDs to keep the anonymity of community members, and \revtext{removed all posts with ``deleted'' or ``removed'' status}. Next, we paraphrased all quoted posts instead of using direct quotes, preventing traceability to the original posts. \revtext{As prior work has cautioned that paraphrased quotes can still be identifiable~\cite{reagle2022disguising}, we conducted this step on an iterative basis, searching the paraphrased quotes using Google and Reddit search~\cite{reagle2022disguising} to assess whether they could still be traced back to their original sources. In our final approach, the paraphrasing not only involves semantic restructuring, but also includes the deliberate blurring or removal of disclosed details~\cite{van2020fluids,fiesler2024remember}. The procedure sacrifices some fidelity yet effectively prevents traceability based on the wording and details, and the rephrased examples remain representative of their categories for readers' understanding. Finally, regarding data management, the collected data was securely stored on a password-protected computer accessible only to the research team, and \revtext{will be permanently deleted after the study}.}
    \item \revtext{This study involves text classification to efficiently identify users' fairness concerns. Such AI use warrants particular attention to prevent potential harms to intellectual property and the privacy of community users when their posts are used in an automated pipeline. To address these concerns, we trained and used the RoBERTa model entirely offline. The local fine-tuned classifier, as a discriminative model, lacks generation or memorization capabilities and will not be released online.}
    \item \revtext{As highlighted by~\citet{fiesler2024remember}, Reddit researchers should holistically consider harms and benefits at a community level, not just at the individual level. In particular, sharing research with the community should be actively (and carefully) considered~\cite{fiesler2024remember,fiesler2018participant}. In our setting, we consider sharing the fair recruitment framework with the community as a beneficial practice that could promote fair hiring and literacy in algorithmic hiring. Therefore, as part of future work, we will share a summarized and accessible version of our framework back with the subreddit while being attentive to community norms, and we will also continuously monitor community users' feedback.}
\end{enumerate}



\section{FINDINGS AND IMPLICATIONS}\label{Finding-Implication}

The qualitative analysis of r/jobs posts revealed a comprehensive taxonomy of job seekers' fairness concerns (RQ1). It covers four overarching themes, including \textit{\revtext{personal attribute discrimination beyond legally protected attributes}} in Section \ref{RQ1-1}, \textit{interaction bias} in Section \ref{RQ1-2}, \textit{improper interpretations of qualifications} in Section \ref{RQ1-3}, and \textit{power imbalance} in Section \ref{RQ1-4}. \modtext{In each section, we include corresponding design implications (RQ2) for online hiring systems from VSD to support their traceability to empirical and conceptual investigation. For each design implication, we present its development through VSD to facilitate interpretation, with \textit{empirical investigation} grounded in r/jobs posts, \textit{conceptual investigation} to unpack normative dimensions of fairness concerns, and \textit{technical investigation} illustrating how algorithmic and interface design can embody or support fairness-related values.}




\subsection{\revtext{Personal Attribute Discrimination Beyond Legally Protected Attributes}}\label{RQ1-1}



Discrimination against individuals based on personal attributes, such as race, gender, sexual orientation, religion, and disability status, is broadly documented in the existing literature on hiring and job markets~\cite{kline2022systemic,purkiss2006implicit,quillian2017meta,riach2002field}. Our qualitative analysis revealed a variety of users' fairness concerns about \revtext{personal attribute discrimination, including both legally protected attributes (e.g., age, gender, disability) and less commonly examined ones (e.g., family and geolocation)}. The fairness concerns, along with the code descriptions and (paraphrased) examples, are presented in Table \ref{tab:bias_taxonomy_sensitive_attri}.

In general, job seekers' fairness concerns about \revtext{personal attribute discrimination} spanned all stages of the hiring pipeline, from sourcing, through CV screening, to candidate evaluation. Notably, in addition to well-documented protected attributes such as race and gender~\cite{suhr2021does,hu2022balancing}, we identified some factors that have been less highlighted in fair recruitment algorithms \revtext{and non-discrimination laws}, although users have expressed extensive concerns about them. For example, job seekers expressed concerns about the influence of \textit{appearance bias} in recruitment, such as how hairstyle, facial hair style, or dress style may convey negative impressions in their CVs, recorded video presentations, and interviews. Moreover, \textit{family bias} has been mentioned by some job seekers as they believed their marriage and children contributed to the job rejections, especially in intersectionality with age and gender (e.g., discrimination against single moms). \textit{Geolocation bias} has also been widely noted as the discrimination against non-local job seekers and candidates from specific geographic areas. 

Our analysis furthermore showed that bias towards \revtext{personal} attributes was not always based on job seekers' direct personal disclosure; rather, these attributes were often indirectly inferred from relevant experiences and characteristics (as proxy attributes~\cite{zhu2023weak}). For instance, posters mentioned the inference of age from education and work background, as well as the inference of race and gender from job seekers' names. Finally, our findings revealed that \revtext{personal} attributes may often intertwine with each other in ways that amplify discrimination. For instance, the appearance bias may affect people of a certain age, race, and gender more strongly than others.

\renewcommand{\arraystretch}{1.5}
\begin{table*}
  \scriptsize
  \caption{Job seekers' fairness concerns related to \revtext{personal attribute discrimination beyond legally protected attributes}.}
  \label{tab:bias_taxonomy_sensitive_attri}
  \begin{tabular}{p{2cm}p{5.3cm}p{5.3cm}}
    \toprule
    Fairness Concern & Description & Example \\
    \midrule
    
    ageism & Bias based on age. Often inferred from education, working experience, appearance, and future plans & \textit{\revtext{Everywhere I apply, they tend to hire younger candidates. It makes me feel like our experience doesn't hold any value.}} \\
    \cline{2-3}

    sexism & Bias based on being perceived as male or female, and relevant stereotypes of gender roles. Can be direct or indirect (e.g., pregnancy and children); Manifested in both outcome fairness and procedural fairness (e.g., additional formalities)  & \textit{\revtext{Would it be unethical to use a gender-neutral name on my resume to avoid potential gender bias?}} \\
    \cline{2-3}

    LGBTQ+ bias & Bias against LGBTQ+ individuals. Often intersect with appearance bias. & \textit{\revtext{Being transgender appeared to influence the hiring decision.}} \\
    \cline{2-3}

    ethnicity bias & Bias based on ethnicity. & \textit{\revtext{Statistics highlight the disadvantages for Black and ethnic minority candidates in job applications.}} \\
    \cline{2-3}
    
    nationality bias & Bias based on nationality. Often inferred from name and language. & \textit{\revtext{Less foreign-sounding names often receive more responses.}} \\
    \cline{2-3}

    appearance bias & Bias based on appearance. Taking effect in diverse stages (e.g., CV and interview); often intersectional (e.g., with ageism and sexism) & \textit{\revtext{some HR professionals may discriminate against unconventional hairstyle, favoring candidates who conform to mainstream appearance norms.}} \\
    \cline{2-3}

    disability/health bias & Bias based on disability and health conditions. Nuanced differences in visible/non-visible and mental/physical conditions. & \textit{\revtext{I was rejected because I'm unvaccinated despite valid reasons.}} \\
    \cline{2-3}

    geolocation bias & Bias based on geolocation, such as discrimination against non-local job seekers and individuals from specific areas. & \textit{\revtext{I'm certain I'm facing discrimination based on my location as I'm getting a lower response rate compared to local opportunities.}} \\
    \cline{2-3} 

    family bias & Bias based on family issues such as marriage, pregnancy, children, etc. & \textit{\revtext{Is it risky to mention being a mom in a job interview?}} \\
    \cline{2-3}

    political bias & Bias based on political affiliation. & \textit{\revtext{I'm concerned that being seen as conservative may be viewed negatively after several rejections.}}\\
    
    \cline{2-3}
    religion bias & Bias based on religious beliefs. & \textit{\revtext{I worry that religion might quietly factor into decisions, even if it's explained as something else.}}\\

    \bottomrule
  \end{tabular}
\end{table*}

\subsubsection{Design Implications}
\begin{figure}
	\centering
		\includegraphics[scale=.37]{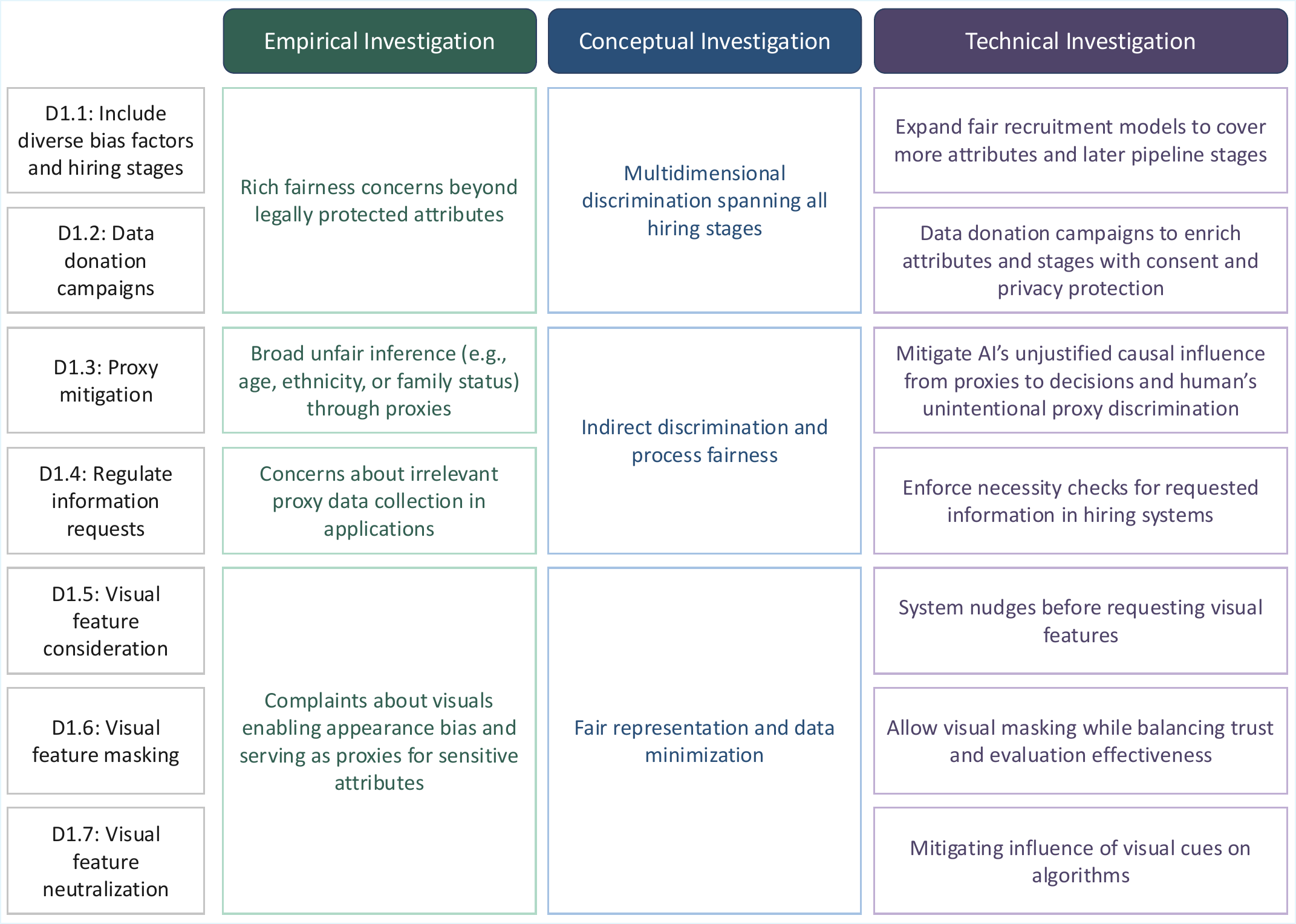}
	\caption{VSD based on ``\revtext{personal attribute discrimination beyond legally protected attributes}'' concerns. The technical investigation in this figure focuses on how design elements can embody or support identified values. Related analytical processes, such as identifying value misalignments in current technology to derive design implications, are detailed in the text.}

	\label{FIG: VSD-1}
\end{figure}

\modtext{
We illustrate the VSD processes, including empirical, conceptual, and technical investigations, based on concerns related to \textit{\revtext{personal attribute discrimination beyond legally protected attributes}} in Figure \ref{FIG: VSD-1}.}

\paragraph{Accounting for more diverse \revtext{personal} attributes at multiple stages of fair recruitment systems}  
Mainstream fair hiring algorithms mostly focus on the sourcing and screening stages (e.g., job recommendation and CV ranking)~\cite{li2021algorithmic}, and model only a small subset of sensitive attributes (e.g., only race and gender)~\cite{fabris2023fairness,suhr2021does,hu2022balancing}. Publicly available datasets for algorithmic fairness research also exhibit corresponding imbalances across sensitive attributes and system stages. For example, a recent survey on fairness in algorithmic hiring revealed that existing datasets covered a narrow set of sensitive variables, primarily focusing on gender and race/ethnicity~\cite{fabris2023fairness}. Nonetheless, we find that users' fairness concerns are multifaceted. Factors such as appearance bias, disability bias, and location bias, though less covered in fair hiring research and datasets, are widely mentioned by job seekers. Moreover, some of these factors primarily affect later stages of the hiring pipeline, such as appearance bias during interviews and health bias in background checks, rather than the sourcing or screening stages. Therefore, we suggest researchers and industry practitioners actively \textbf{include diverse bias factors and hiring stages (D1.1)\footnote{We use ``D'' to denote Design insights/considerations.}} when building fair recruitment systems. In particular, when existing datasets cover limited features to train and evaluate fair hiring algorithms, there is great potential to organize \textbf{data donation campaigns (D1.2)} with informed consent~\cite{gomez2023beyond} to establish datasets with expanded dimensions of \revtext{personal} attributes for fair hiring research. Such efforts, however, must navigate practical challenges including privacy protection, ethical data governance, and ensuring the representativeness of collected data~\cite{he2026when}.


\paragraph{Mitigation for sensitive attribute proxies within fair recruitment systems}
The findings highlight the \textit{implicitness} of sensitive attributes with broad proxy characteristics existing in the hiring process. Typical examples include inferring age from education and working background, inferring ethnicity from the name, inferring geolocation from contact information, and inferring family status from social media background checks. \revtext{Such empirical findings mirror real-world cases, such as Amazon's discontinued AI recruiting system that was shown to disadvantage graduates of women's colleges as a gender proxy~\cite{WomenCollegeDis,bornstein2018antidiscriminatory}.} Besides, the implicitness of sensitive attributes underscores the vulnerability of some pre-processing approaches for fair hiring algorithms that mainly consider explicit variables. For example, both rule-based scraping~\cite{de2019bias} (automatically removing words related to sensitive attributes) and rule-based substitution~\cite{rus2022closing} (automatically neutralizing all words related to sensitive attributes) may fail to produce a debiased resume - \revtext{in other words, ``fairness through unawareness'' is generally ineffective.} Therefore, designers of hiring algorithms and systems are suggested to critically reflect on \textbf{proxy mitigation (D1.3)}, i.e., how to avoid commonly believed non-sensitive qualifications from being leveraged as proxies for protected sensitive attributes. \revtext{For AI developers, rather than eliminating or neutralizing all proxies, a more practical strategy is to block unjustified causal influence from proxies to decisions~\cite{parraga2025fairness}. In particular, recent responsible AI research has proposed detecting proxies by probing whether sensitive attributes can be predicted from features or embeddings~\cite{parraga2025fairness}, and then mitigating such influence via methods such as adversarial debiasing~\cite{zhang2018mitigating} or path-specific fairness constraints~\cite{chiappa2019path,plecko2025fairness}. For HCI practitioners, it is crucial to design and evaluate various interventions and nudges for decision-makers to enhance awareness of proxy influence and mitigate unintentional proxy discrimination, such as disclosing proxy correlations~\cite{goyal2024impact} or providing fair machine guidance~\cite{yang2024fair}.}



Besides, from a policy standpoint, we also suggest that \textbf{regulators closely monitor and enforce the limits on what information may be requested from candidates (D1.4)} in hiring systems as a basic approach to alleviate proxy-based discrimination. In particular, before deploying a hiring system, it is necessary to assess whether each piece of requested information is meaningfully relevant to qualification assessment, or potentially functions as a proxy for protected attributes. Indeed, such protections could be derived from the legal principles of \textit{indirect discrimination} in European non-discrimination law~\cite{EuroLaw} and \textit{disparate impact} in US non-discrimination law~\cite{USLaw}, which apply when a practice appears neutral but ultimately discriminates against disadvantaged groups.

\paragraph{Handling of visual data} 
The findings suggest that some user-perceived biases are more visually related (e.g., appearance and disability bias) based on photos, recorded videos, or synchronous interviews, which challenges the practicality of traditional debiasing approaches primarily relying on text and tabular data. Therefore, it is crucial for researchers and developers of fair hiring systems to critically reflect on debiasing in multimodal settings. In particular, we propose a standardized pipeline for processing visual and video data, while recognizing that such proposals remain largely conceptual and require rigorous technical and ethical evaluations. First, as photos and videos often contain visual cues that function as proxies, hiring systems might incorporate a nudge interface prompting employers to deliberate on the necessity (i.e., \textbf{visual feature consideration, D1.5}): do you really need to collect personal photos or recorded videos from candidates? Second, if visual and video data are considered necessary, hiring systems can explore the use of virtual avatar systems for videos or synchronous interviews to obscure proxy indicators (i.e., \textbf{visual feature masking, D1.6})~\cite{crone2022interview,behrend2012effects}. Further, in contexts where avatars are infeasible, hiring systems may offer preprocessing methods to mitigate visual-based \revtext{personal attribute discrimination} (i.e., \textbf{visual feature neutralization, D1.7}), such as using face decorrelation algorithms~\cite{hemamou2021don}. Nonetheless, future research should examine the trade-offs between the effectiveness of visual debiasing and evaluation validity (e.g., its impact on trust, perceived authenticity, and user comfort).


\subsection{Interaction bias}\label{RQ1-2}

We identified job seekers' nuanced fairness concerns during interactions with hiring systems and companies, as shown in Table \ref{tab:bias_taxonomy_process}. They span applicant–system and applicant–recruiter interactions throughout the hiring lifecycle, including initial outreach, applications, interview coordination, and post-interview communication. Job seekers believed that recruiters were very likely to form biased impressions during the interaction and, intentionally or unintentionally, make hiring decisions based on them. These fairness concerns are primarily related to the level of \textit{individual fairness} in hiring~\cite{hauer2021legal}, i.e., ensuring that similarly qualified candidates receive comparable outcomes without discrimination based on irrelevant traits. Compared to group fairness, individual fairness has gained much less attention from researchers and developers for (algorithmic) hiring systems~\cite{hauer2021legal}.

\begin{table*}
  \scriptsize
  \caption{Job seekers' fairness concerns related to interaction bias.}
   
  \label{tab:bias_taxonomy_process}
  \begin{tabular}{p{2cm}p{5.3cm}p{5.3cm}}
    \toprule
    Fairness Concern & Description & Example \\
    \midrule
    communication method & Bias based on how job applicants connect with the company, e.g., mail vs. email, private channels vs. public channels, company websites vs. recruitment tools, etc. & \textit{\revtext{I've only ever been hired through paper applications. Online applications always seem to go nowhere.}} \\
    \cline{2-3}

    self-presentation approach & Bias based on how job applicants present themselves, e.g., the structure and font of their resume, the design of their personal website, etc. & \textit{\revtext{I was rejected everywhere. I’m not sure if my résumé design made things worse.}} \\
    \cline{2-3}
    
    interview process & Bias based on how the interview happens, e.g., the time, interviewing order, availability, and the time difference of the interview & \textit{\revtext{Hiring managers tend to be more attentive early on, which can make the first candidate more memorable.}} \\
    \cline{2-3}

    evaluation method & Bias based on how evaluation works, e.g., unsuitable evaluation systems and unreasonable interview questions & \textit{\revtext{I took a pre-employment assessment. It included discriminatory questions that conveyed a classist tone.}}\\
    \cline{2-3}

    technical issue & Limitations of the technical systems that may bring unequal outcomes, e.g., universities not in the database & \textit{\revtext{I excel in phone interviews and am often highly qualified, yet I lose opportunities because of a flawed system.}}\\
    \cline{2-3}

    opt-out decision & Opt-out decisions that may influence impression, e.g., opt-out AI evaluation and sensitive identity disclosure & \textit{\revtext{I know this is voluntary, but could choosing not to answer sensitive questions affect my application?}}\\
    \cline{2-3}

    questions and feedback & Bias based on whether and how job seekers raise questions before/during/after the interview & \textit{\revtext{Before interview I asked about the technical format and now worry that it might seem seeking an unfair advantage and hurt my chances.}}\\

    \bottomrule
  \end{tabular}
\end{table*}

\subsubsection{Design Implications}

\begin{figure}
	\centering
		\includegraphics[scale=.37]{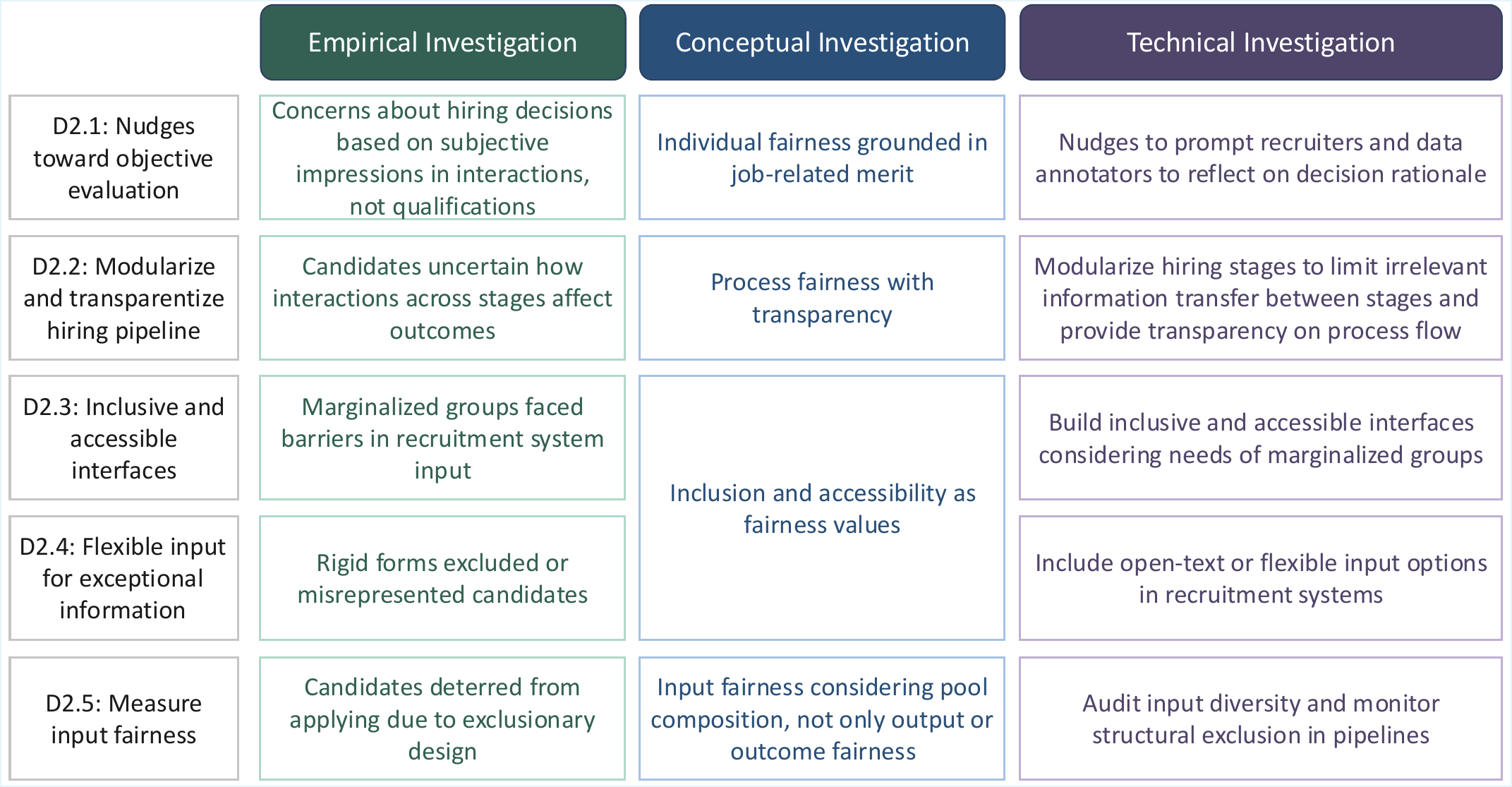}
	\caption{\modtext{VSD based on ``interaction bias'' concerns.}}
	\label{FIG: VSD-2}
\end{figure}

\modtext{
The VSD processes according to \textit{interaction bias} are shown in Figure \ref{FIG: VSD-2}.}

\paragraph{Nudges toward objective evaluation} The interaction bias highlights a significant challenge in the hiring decision-making process: recruiters' evaluations might be influenced by recruiter-candidate interactions, which can sometimes deviate from objectively measuring individual qualifications. For example, candidates' communication styles or self-presentation strategies may not align with recruiters' subjective preferences, potentially introducing bias into hiring decisions. To this end, it is warranted to develop and examine design approaches that support recruiters in making impartial decisions, minimally influenced by irrelevant individual interactions. In particular, \textbf{nudges toward objective evaluation (D2.1)}, which prompt recruiters to reflect on the rationale behind candidate selection or rejection, may help foster more rational and transparent hiring. Such nudges can also be enhanced with AI, for example, by generating personalized prompts that highlight inconsistencies or irrelevant evaluation patterns~\cite{yang2024fair}. Notably, nudges toward objective evaluation may not only apply to real-world recruiters' decision-making, but also be a potentially valuable practice for data annotation. For example, nudging annotators to score CVs more based on qualifications rather than being influenced by CV structure and font might be helpful in creating a higher-quality dataset to build fair hiring algorithms.

\paragraph{Standardization of the hiring pipeline to increase the perceived process fairness} Job seekers' concerns about interaction bias reflect how recruiter-candidate interactions can be perceived as inaccurate proxies for evaluating qualifications, introducing opacity into the hiring process. Notably, job seekers typically did not (and could not) provide direct evidence of interaction bias. Instead, their concerns often manifested as preconceived assumptions (i.e., that certain interaction styles would be penalized) or post hoc attributions (i.e., explaining failure through perceived interpersonal missteps). In other words, the perceived uncertainty in how interactions influence decisions poses barriers to users' trust in hiring fairness. This insight points to a promising design direction: standardizing the hiring pipeline to reduce ambiguity and increase perceived fairness. Specifically, we highlight two critical components: \textbf{modularization and transparentization of the hiring pipeline (D2.2)}. First, modularizing the hiring pipeline could play a pivotal role in minimizing the transfer of irrelevant information across hiring stages, reducing the influence of unrelated interactions on decision-making. For instance, separating the modules of CV collection and candidate evaluation could reduce the impact of communication channels on the candidate selection process. Additionally, adopting a module for CV parsing and anonymizing before qualification assessment could reduce the influence of CV structure and design if it is irrelevant to employment duties. Following modularization, increasing transparency, such as informing candidates how decisions are staged and which stakeholders are involved, can further improve perceived fairness. For example, informing candidates that the selection of interview time and opt-out decisions would not be visible to recruitment evaluators in the hiring system might be beneficial in enhancing the perceived fairness of the process for job seekers. Nonetheless, the implementation of modular and transparent hiring systems may vary across organizations and must take into account technical feasibility and organizational structure.

\paragraph{Inclusive and accessible interfaces for \revtext{marginalized groups}}
Another important aspect of interaction bias involves how certain system interactions intersect with job seekers' backgrounds, potentially compounding disadvantages for marginalized groups. For instance, individuals with educational experience in the Global South may encounter technical barriers, such as their universities not being listed in system databases. Similarly, candidates with speech impairments may find fixed-time self-presentation formats inadequate to represent their capabilities. To this end, we suggest \textbf{inclusive and accessible interfaces for \revtext{marginalized groups} (D2.3)}. As a foundational step, allowing flexible input mechanisms can support inclusion in computer-mediated hiring systems. Hiring systems are encouraged to support \modtext{\textbf{flexible input for exceptional information (D2.4)}}, such as open text fields for educational or employment background, especially when standard options may exclude or misrepresent certain groups. Beyond basic inclusion, designers should proactively identify and address nuanced interaction barriers faced by specific marginalized groups, and develop specialized interface adaptations to enhance accessibility. For example, with advances in accessible video conferencing~\cite{rui2022online,leporini2023video}, HCI researchers and practitioners are encouraged to embed relevant design guidelines into specialized video interview platforms tailored for underrepresented groups (e.g., users with hearing, speech, or visual impairments). 



\paragraph{Measuring input fairness} \revtext{Current fairness measures in hiring mostly focus on fairness after applicant data enters the hiring system (e.g., rankings and acceptance rates across different applicant groups), while often overlooking structural biases in data composition as part of a system-wide scrutiny~\cite{alexander2025sourcing,fabris2023fairness}.} In fact, interaction bias signals that certain job seekers may be discouraged or excluded from applying because of interface designs that fail to accommodate their needs. It is imperative to account for and address the underrepresentation of protected identities in applicant pools, which often stems from structural barriers or exclusionary practices. \textbf{Measuring input fairness (D2.5)} as part of a system evaluation, i.e., analyzing the composition of applicant pools, can surface problematic aspects of the hiring system more broadly, for instance, by discouraging non-inclusive application interfaces.

\subsection{Improper interpretations of qualifications}\label{RQ1-3}

Table \ref{tab:bias_taxonomy_sensitive_quali} demonstrates how job seekers raised fairness concerns about improper interpretations of their qualifications in the hiring process. It covers various non-protected factors, including \textit{education}, \textit{working history}, \textit{references}, \textit{personality}, and \textit{\revtext{negative background history}}. Job seekers' concerns about the fairness of interpreting such qualifications reflect two levels of tension. First, some job seekers perceived that the interpretation of these qualifications could reinforce social disparities. For instance, social stratification could result in variations in educational background and the strength of referral letters for individuals with similar skills; discrimination in the workplace might lead to employment gaps that are difficult to explain to future employers; marginalized groups might experience unfair treatment in social life, which can disadvantage them during background checks. In addition, job seekers expressed that some of these factors may have a limited correlation with actual job-related skills, yet still influence hiring outcomes due to their symbolic or socially coded meanings. These reflections highlight how standard qualifications, although seemingly neutral, can sometimes reflect or perpetuate structural inequalities. Such interpretive patterns illustrate how hiring systems can inadvertently contribute to the social construction of inequality~\cite{ore2000social}.

\begin{table*}
  \scriptsize
  \caption{Job seekers' fairness concerns related to improper interpretations of qualifications.}
  \label{tab:bias_taxonomy_sensitive_quali}
  \begin{tabular}{p{2cm}p{5.3cm}p{5.3cm}}
    \toprule
    Fairness Concern & Description & Example \\
    \midrule
    education & The education level, region, school, and department may inherit social inequality and may not well correspond to the necessary skills required for a job. & \textit{\revtext{I have a degree from a third-world country that meets Western education standards, but employers seem incredibly biased.}} \\
    \cline{2-3}

    working history & Working history may be influenced by personal decisions (e.g., health conditions leading to short-term work and job gaps) and working environment (e.g., discrimination in the workplace leading to employment gaps), both correlated with social inequality. & \textit{\revtext{I left a role after a short tenure due to a poor environment and now worry about how this history may be judged if asked.}} \\
    \cline{2-3}

    references & References/recommendation letters reflect individuals' social capital, which might inherit social inequality, workplace prejudice, and disputes over interests. & \textit{\revtext{I worry that my current employer might respond negatively in the references as my leaving would inconvenience them.}} \\
    \cline{2-3} 

    personality & Personality (e.g., being an introvert or an extrovert), often shaped by the social environment, may influence hiring decision-making but not well represent qualifications. & \textit{\revtext{After a long job search, I feel being introverted puts me at a disadvantage.}} \\ 
    \cline{2-3} 

    \revtext{negative background history} & \revtext{Negative background history} (e.g., criminal record), fundamentally influencing background checks, may inherit societal prejudice.  & \textit{\revtext{I belong to a racial minority and had false charges that were later dismissed. I worry they could still be used to discriminate against me.}} \\

    \bottomrule
  \end{tabular}
\end{table*}

\subsubsection{Design Implications}

\begin{figure}
	\centering
		\includegraphics[scale=.37]{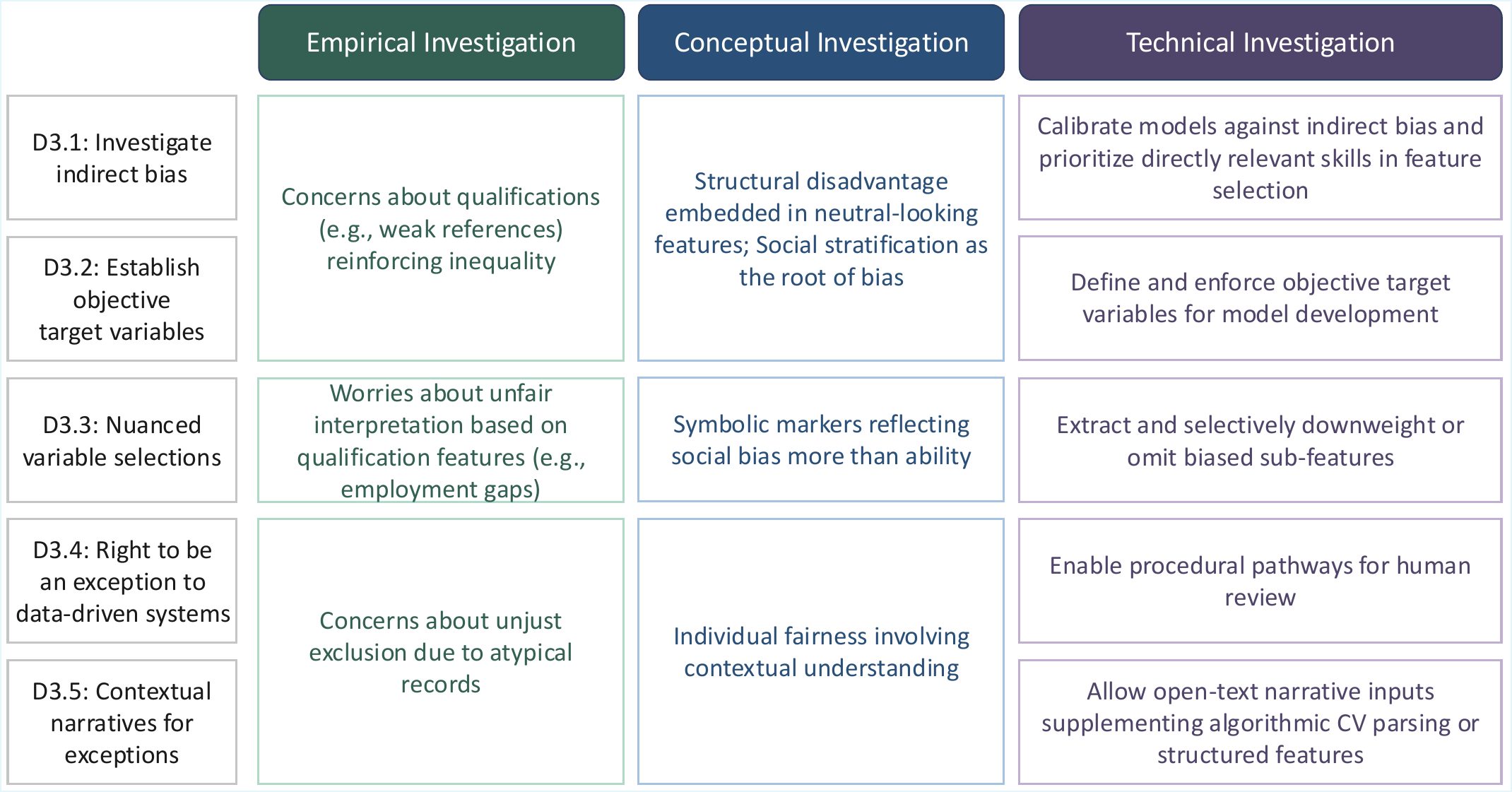}
	\caption{\modtext{VSD based on ``improper interpretations of qualifications'' concerns.}}
	\label{FIG: VSD-3}
\end{figure}

\modtext{
The VSD processes according to \textit{improper interpretations of qualifications} are illustrated in Figure \ref{FIG: VSD-3}.}

\paragraph{Investigating indirect bias} The findings signify that institutional biases can be inherited, accumulated, and reinforced across life stages. For example, workplace discrimination may result in early job departure and weak reference letters, which can in turn lead to job search difficulties and employment gaps, creating a cycle of disadvantage over time. Although less visible than direct discrimination based on protected attributes, such indirect bias poses a significant challenge to hiring systems and is difficult to eliminate. Unlike protected attributes such as race and gender, qualifications like education and work history cannot be easily removed or neutralized, as they play a key role in candidate evaluation. Some scholars have recognized that underlying group-based inequalities embedded in qualifications contribute to indirect bias, and have proposed corresponding mitigation strategies. For example, Booth et al. proposed group normalization as a bias mitigation strategy, dividing job seekers' data based on their sensitive group membership and normalizing each feature within the group~\cite{booth2021bias}. Nonetheless, this method may overlook the intersectionality of sensitive attributes and still allow for intra-group disparities~\cite{booth2021bias}. Therefore, we call for further research into \textbf{investigating indirect bias (D3.1)} as part of developing fair ranking algorithms. One potential direction is to prioritize directly relevant skills and abilities (e.g., programming expertise), while cautiously reducing reliance on indirect indicators (e.g., educational background) when appropriate. In addition, algorithm developers may incorporate calibration methods that account for the correlation between protected attributes and qualification variables, to help mitigate accumulated (dis)advantages. Moreover, indirect bias underscores the importance of \textbf{establishing objective target variables (D3.2)} during model development, for instance, prioritizing task-relevant skills and experience over proxies such as similarity to previous hires~\cite{fabris2023fairness}.

\paragraph{Careful variable selections and feature interpretations} The findings indicate that seemingly appropriate qualification features may contain components that unintentionally introduce bias. For example, while work experience is generally considered a neutral attribute, specific components, such as employment gaps, may trigger biased interpretations. Similarly, educational background provides essential context about a candidate, but elements like university ranking or field of study may activate subjective or stereotypical judgments in certain contexts. Therefore, we suggest performing \textbf{nuanced variable selections (D3.3)} to facilitate fair recruitment. To support this, we recommend developing preprocessing components that extract fine-grained sub-features, enabling selective anonymization or adjustment of those that are contextually sensitive. For example, structured extraction techniques can be used to identify key elements in CVs, such as job title, company name, duration, and responsibilities. These extracted variables can then be assessed to determine their relevance and potential impact on fairness, with the option to de-emphasize or omit those considered potentially biased, such as employment gaps, depending on the context. However, such fine-grained selection also requires clear fairness criteria and domain-specific justification to avoid unintended information loss or fairness–accuracy trade-offs.

\paragraph{Supporting the right to be \modtext{an exception to data-driven rules}.} Job seekers' concerns about \revtext{negative background history} suggest that societal prejudice can contribute to the creation or amplification of adverse records, such as potentially inaccurate criminal charges, that have long-term impacts and severely limit employment opportunities. Similar patterns can be found in lost educational opportunities or prolonged employment gaps resulting from institutional biases. Hiring algorithms risk perpetuating such structural disadvantages if they incorporate background features without contextual understanding. To this end, we advocate for recognizing \textbf{\modtext{the right to be an exception to data-driven rules (D3.4)}}~\cite{cen2023right} in algorithmic hiring, a principle that supports individualized consideration when automated models risk unfair exclusion. Hiring systems should provide procedural space for such exceptions by allowing individualized review in cases where automated decisions may be unduly influenced by biased data. More broadly, we recommend fostering human-algorithm collaboration that enables discretion, review, and contextual interpretation, rather than enforcing a rigid, sequential pipeline. For example, systems may allow candidates to provide human-authored open-text inputs as \textbf{contextual narratives for exceptions (D3.5)}, which can serve as supplemental inputs for automated CV evaluation, akin to the use of natural language prompts in recommendation systems~\cite{radlinski2022natural}. Such narratives can enable candidates to explain or contextualize atypical records, potentially mitigating the impact of structural disadvantage. Nonetheless, it is necessary to examine potential backfire effects (e.g., how people might exploit the system) and develop intervention approaches (e.g., moderators that assess the credibility of contextual narratives).

\subsection{Power Imbalance}\label{RQ1-4}

Last but not least, power imbalance has emerged as a significant concern for job seekers regarding fairness (see Table \ref{tab:bias_taxonomy_sensitive_power}). It includes three major topics: (1) \textit{Power imbalance between job seekers and hiring platforms}. Job seekers noted that recruitment platforms, as the information hub centralizing recruitment-related resources, may sometimes disadvantage (rather than facilitate) job seekers, such as promoting unrelated job recommendations and leaking job application information to unrelated companies; (2) \textit{Power imbalance between job seekers and companies}. Some job seekers expressed concerns about the power asymmetry between job seekers and companies \modtext{given} the inherent imbalance of job demand and supply. The opacity of the hiring process significantly reinforces perceptions of this power imbalance. For instance, job seekers widely complained about suddenly losing all communication from the company (ghosting) or receiving rejections without any reason; (3) \textit{Power imbalance between different job-seeking communities}. The imbalanced representation of different job-seeking communities leads to information asymmetry. As a result, online knowledge and experience related to job-seeking are often inapplicable to specific populations. Users' concerns encompass a broad range of representation biases, including identity (majority vs. minority), occupation (white-collar vs. blue-collar jobs), and experience (senior vs. entry-level positions). These factors present another source of inequality from a macro perspective.

\begin{table*}
  \scriptsize
  \caption{Job seekers' fairness concerns related to power imbalance.}
  \label{tab:bias_taxonomy_sensitive_power}
  \begin{tabular}{p{2cm}p{5.3cm}p{5.3cm}}
    \toprule
    Fairness Concern & Description & Example \\
    \midrule
    
    power imbalance between job seekers and hiring platforms & Power asymmetry between job seekers and hiring platforms, materializing in ways such as overwhelming ads, unrelated recommendations, information leakage, nuisance calls, etc.  & \textit{\revtext{I keep seeing the same roles in job search because they are ``promoted''.}} \\
    \cline{2-3}

    power imbalance between job seekers and companies & Power asymmetry between job seekers and companies, materializing in ways such as ghosting, inadequate rejection reasons, etc. & \textit{\revtext{I got a delayed rejection much later than promised and received vague feedback on my portfolio that didn't align with the role, which felt frustrating and inconsistent.}} \\
    \cline{2-3} 

    power imbalance between different job-seeking communities & Imbalanced power and information distribution among different communities due to representation bias in rating websites, blogs, etc. & \textit{\revtext{Most job-seeking advice online seems geared toward already advantaged groups in high-salary tech roles.}} \\

    \bottomrule
  \end{tabular}
\end{table*}

\subsubsection{Design Implications}
\begin{figure}
	\centering
		\includegraphics[scale=.37]{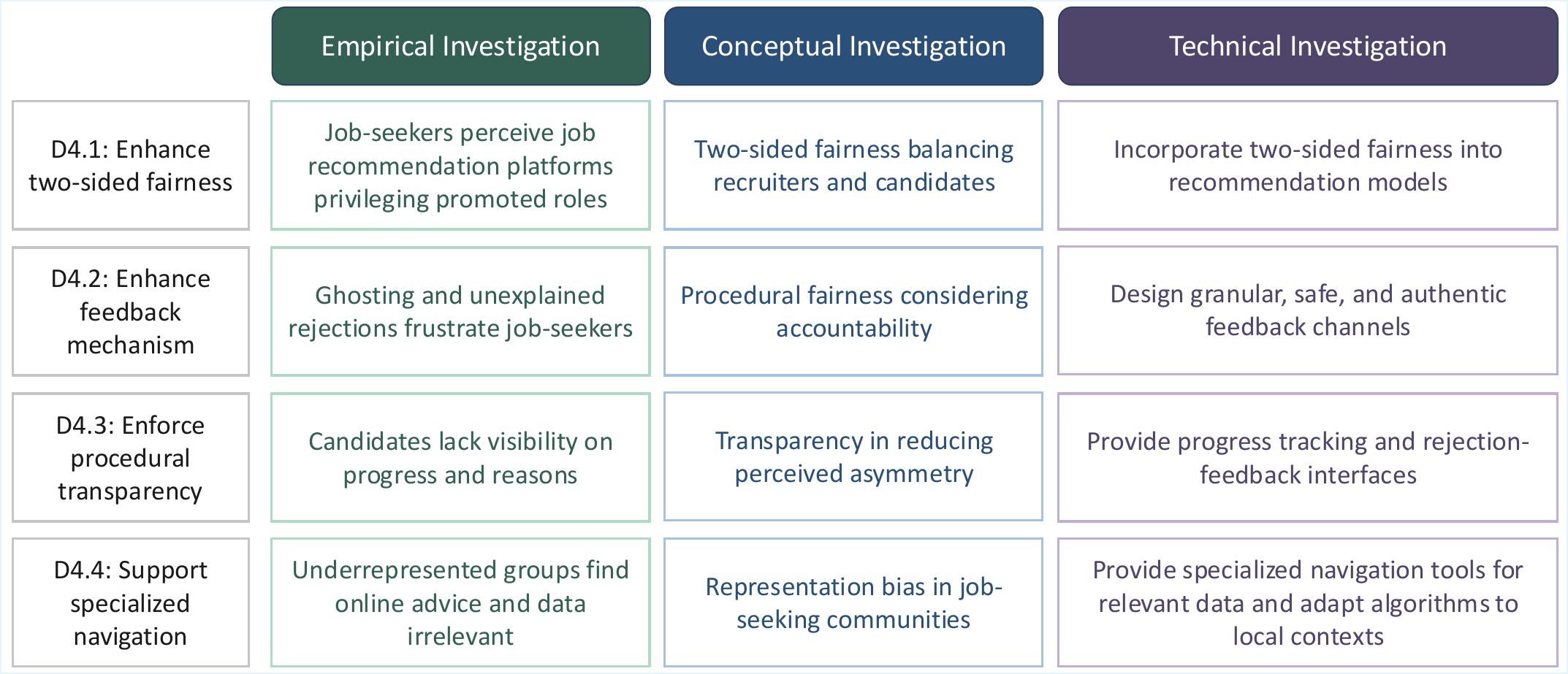}
	\caption{\modtext{VSD based on ``power imbalance'' concerns.}}
	\label{FIG: VSD-4}
\end{figure}

\modtext{
The VSD processes according to \textit{power imbalance} are illustrated in Figure \ref{FIG: VSD-4}.}

\paragraph{Ensuring two-sided fairness in job recommendation.} The findings indicate that the power imbalance between job seekers and hiring platforms might harm user experience and lead to job seekers' disengagement with the platform. Particularly, job seekers perceived that recruitment platforms may disadvantage them through practices such as promoting irrelevant job recommendations or maintaining unclear information-sharing mechanisms. On this note, we suggest \textbf{enhancing two-sided fairness (D4.1)}~\cite{patro2020fairrec,suhr2019two} to balance the needs of both job seekers and hiring companies when developing recommendation models. Two-sided fairness~\cite{patro2020fairrec,suhr2019two} suggests considering both user and item fairness simultaneously, thereby balancing competing interests and promoting justice between different parties. Specific to the job recommendation setting, two-sided fairness ensures not only fair recommendation of job seekers to recruiters, but also fair recommendation of positions to job seekers. In fact, job recommendation algorithms align with the producer-consumer model, similar to most recommendation systems~\cite{patro2020fairrec}. However, unlike most recommendation systems that rely on consumers as the primary source of revenue (e.g., ride-hailing and food delivery), hiring companies as job-post producers can exert influence over promotions and impact the outcomes of job recommendations, making hiring systems susceptible to producer-centered design. We recommend that researchers carefully consider such context-specific incentive structures and power asymmetries when tailoring two-sided fairness models for the hiring context.

\paragraph{Enhancing feedback mechanisms for the hiring process.} The supply and demand relationship of recruitment creates an intrinsic power imbalance between job seekers and hiring companies. Our findings reveal that such power imbalance manifests in diverse ways. It often puts job seekers in a passive and disadvantaged position from the perspective of hiring procedure (e.g., suddenly canceled interviews), information flow (e.g., rejection without reasons), and communication breakdown (e.g., ghosting). On this note, we suggest \textbf{enhancing the feedback mechanism (D4.2)} for the hiring process as a regulating method. While hiring tools have increasingly included the company review and rating as a significant component to engage and inform job seekers (e.g., glassdoor\footnote{https://www.glassdoor.com/} and Kununu\footnote{https://www.kununu.com/}), they largely focus on the employment experience, such as company culture and compensation, while sometimes overlooking the hiring process. Besides, the company review and rating can be utilized to propagate the reputations of employers~\cite{marinescu2021incentives}, as employees' voices tend to dominate over rejected applicants. It is a promising direction to develop and evaluate feedback mechanisms in job-seeking tools that afford granularity (e.g., distinguishing between hiring stages and AI/human stakeholders), safety (e.g., allowing anonymity), and authenticity (e.g., moderating fake reviews). 

\paragraph{Establishing and enforcing basic procedural transparency.} Our findings reveal that the power imbalance between recruiters and candidates often manifests through information asymmetry. For example, candidates perceive unfair treatment when they fail to track the progress of a job application or are not informed of the reason for rejection. Therefore, we call for the collaboration of policy-makers and recruitment systems to \textbf{establish and enforce at least a basic level of procedural transparency (D4.3)} as a countermeasure against the intrinsic recruiter-candidate power imbalance. For instance, it is essential for recruitment systems to offer a progress-tracking interface with a transparent time limit for each stage to ensure timely updates (rather than ghosting), and provide a rejection-feedback interface to communicate the reasons for rejection (instead of uninformed rejections).

\paragraph{Reconsidering the generalizability of fair hiring.} The power imbalance between different job-seeking communities reflects the representation bias in the recruitment ecosystem: most of the job search experience sharing centers on only a limited group of professions, cultures, and regions. Such representation bias has a direct impact on the applicability of job-seeking-related knowledge to underrepresented groups. Therefore, it sheds light on the potential for constructing an inclusive and customized information ecosystem for job seekers. First, \textbf{supporting specialized navigation (D4.4)} might be a promising avenue to facilitate communication among specific groups in the information ecosystem, e.g., easing the navigation of experience sharing for underrepresented occupations through diversified hashtags and targeted recommendation mechanisms. More broadly, due to representation bias, the development of fair recruitment algorithms is based on narrow data and applied to limited job and identity categories. For example, fair recruitment algorithms may fail to adequately address (and often overlook) blue-collar jobs in the Global South. To this end, future research and practice should critically assess the contextual transferability of fair recruitment algorithms, ensuring that fairness interventions are tailored to local labor markets, job types, and socio-cultural conditions.

\section{A CONCEPTUAL FRAMEWORK FOR FAIR HIRING}

We conclude by outlining a conceptual framework for fair hiring, developed from a \modtext{VSD} approach. Rather than prescribing fixed solutions, this framework aims to offer a guiding structure to support practitioners in addressing fairness concerns in recruitment systems through both \textit{algorithm} and \textit{interface} design.

\begin{figure}
	\centering
		\includegraphics[scale=.42]{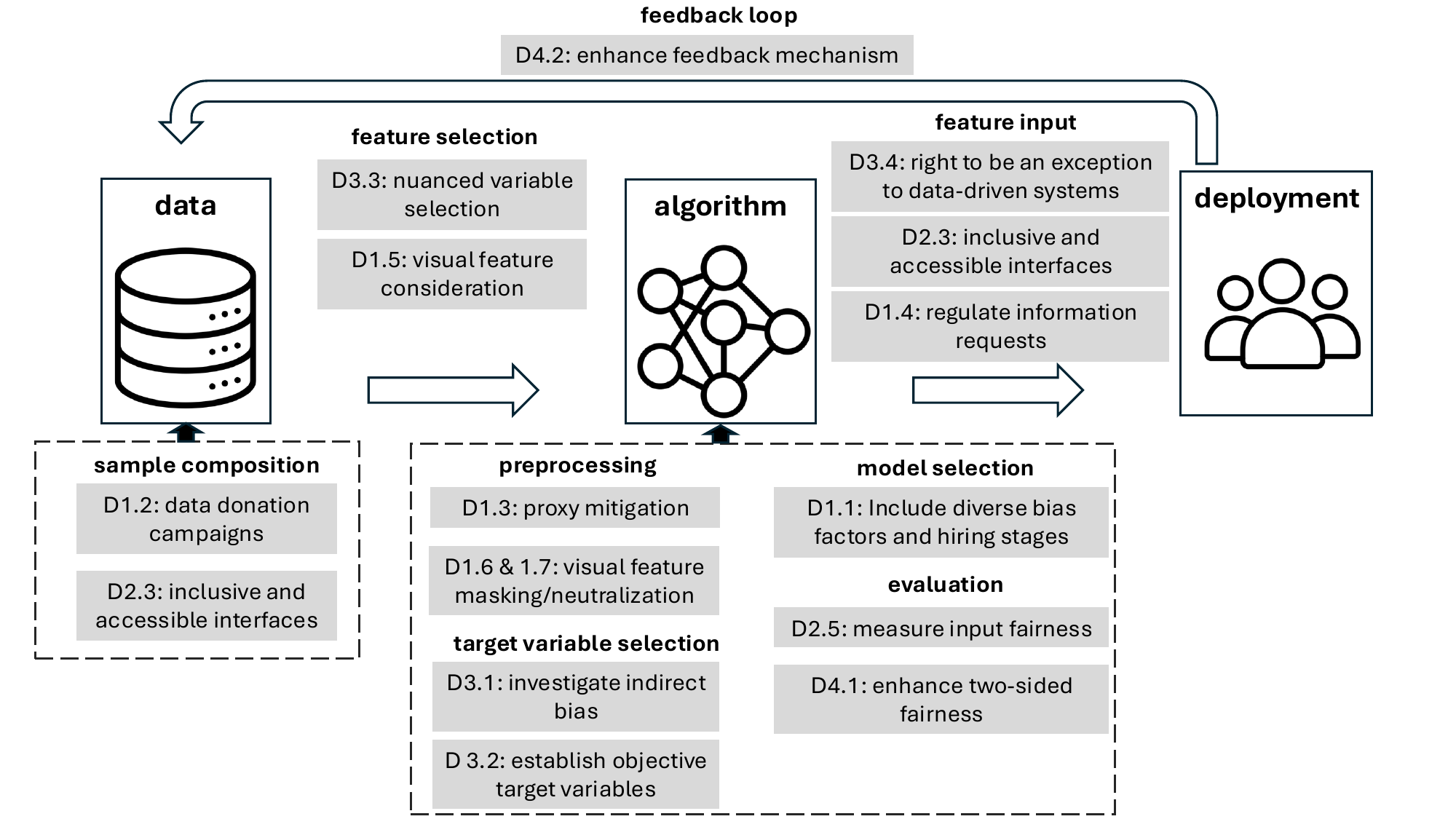}
	\caption{A conceptual framework for designing fair recruitment algorithms, spanning three core phases in the lifecycle of recruitment algorithms: data preparation, algorithm development, and model deployment.}
        \Description{This figure illustrates the framework for designing fair recruitment algorithms. It considers broad algorithmic design stages during the iterative process of data preparation, algorithm development, and model deployment. We connect our design insights with these algorithmic design stages, such as incorporating data donation campaigns for sample composition, nuanced variable selection for feature selection, proxy mitigation for preprocessing, and measuring input fairness for model evaluation. We exemplify this framework with the following case study.}
	\label{FIG: fair_algorithm}
\end{figure}

Figure \ref{FIG: fair_algorithm} illustrates the conceptual framework for designing fair recruitment algorithms. The framework delineates three core phases in the lifecycle of recruitment algorithms, including data preparation, algorithm development, and model deployment~\cite{fabris2022algorithmic}. In particular, we propose data donation campaigns and accessible interfaces to support representative and inclusive \textit{sample composition}. Moving from data to algorithm, \textit{feature selection} emphasizes the nuanced selection of variables and the careful consideration of visual features to ensure fairness before developing algorithms. The implications for the algorithm development stage include key elements covering \textit{preprocessing} (e.g., proxy mitigation), \textit{target variable selection} (e.g., establish objective target variables), \textit{model selection} (e.g., fair algorithms across hiring stages), and \textit{fairness evaluation} (e.g., measure input fairness). Moving from algorithm development to model deployment, \textit{feature input} highlights design interventions to ensure fair and interpretable data processing in real life, including \modtext{the right to be an exception to data-driven rules}, accessible interfaces, and regulated information requests. Finally, the feedback mechanism links challenges that occur during the deployment stage back to data and algorithm design, supporting the continuous refinement of the recruitment system.


\begin{figure}
	\centering
		\includegraphics[scale=.4]{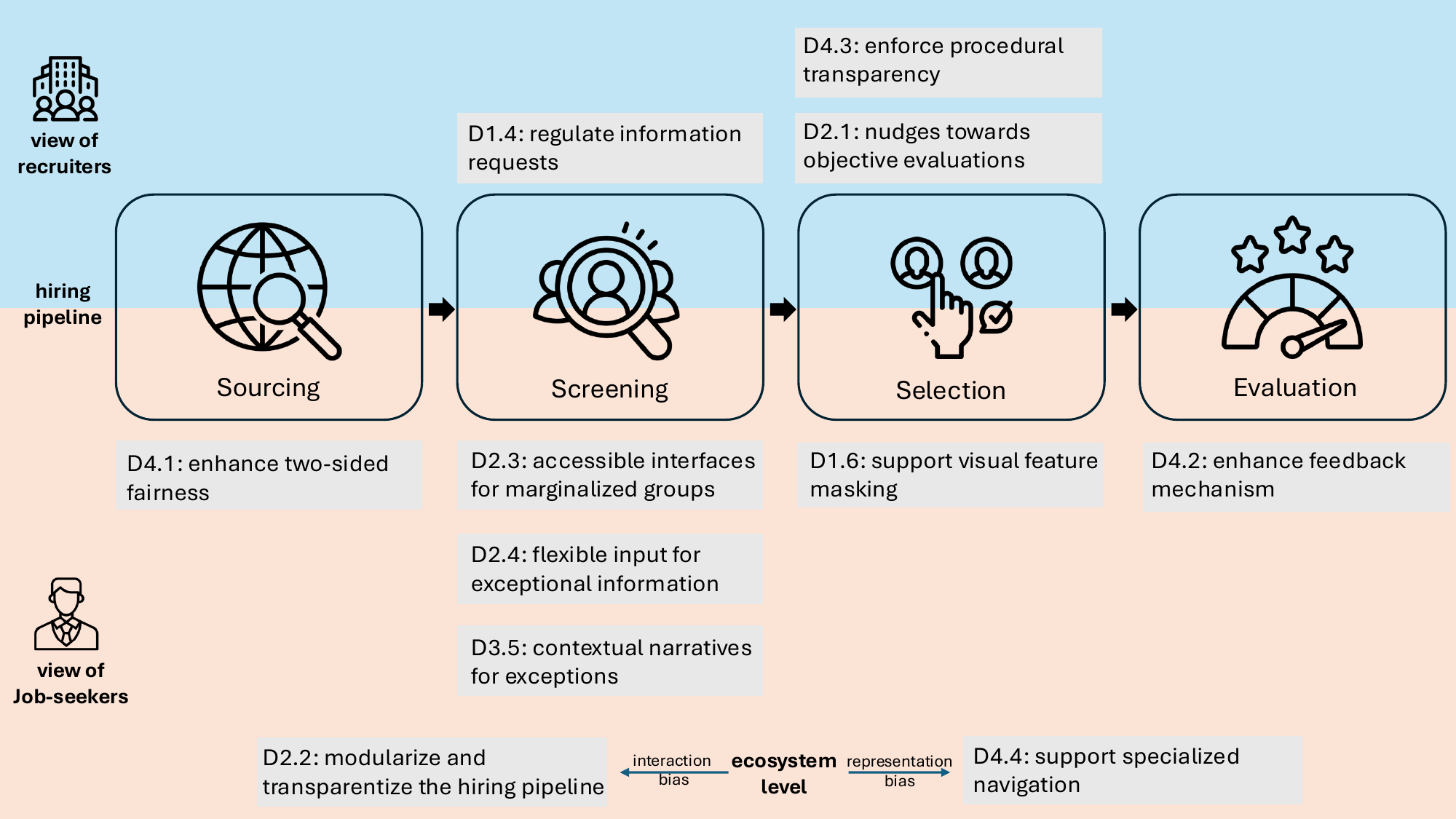}
	\caption{A conceptual framework for designing fair recruitment interfaces, covering four core stages of the online hiring process: sourcing, screening, selection, and evaluation.}
        \Description{This figure illustrates the framework for designing fair recruitment interfaces. We contextualize our design insights within different stages in online hiring, including sourcing, screening, selection, and evaluation, and also consider design elements at the ecosystem level of online hiring systems. We exemplify this framework with the following case study.}

	\label{FIG: fair_system}
\end{figure}

Figure \ref{FIG: fair_system} illustrates the conceptual framework for designing fair recruitment interfaces. The framework is structured along the four core stages of the online hiring process, including \textit{sourcing}, \textit{screening}, \textit{selection}, and \textit{evaluation}~\cite{fabris2023fairness}. We contextualize our design insights within these pipeline stages, while also specifying whether each design element targets recruiters or job seekers. Specifically, during the sourcing stage, we propose enhancing two-sided fairness and generalizing fairness to diverse sourcing methods, thereby promoting fair matching between jobs and job seekers. During screening, we call for accessible interfaces for \revtext{marginalized groups}, flexible input for exceptional cases, and providing contextual narratives for exceptions to mitigate unfair treatments in application processing, and suggest regulating information requests from the recruiters' side. In the selection stage, we encourage objective evaluations via interface nudges to assist recruiters in making fair decisions, and suggest the enforcement of basic transparency in hiring results. For evaluation, the framework focuses on strengthening feedback mechanisms to improve system accountability. At the ecosystem level, we propose modularizing the hiring pipeline to reduce interaction bias, and supporting specialized navigation (e.g., occupation-specific information channels) to facilitate information seeking for underrepresented groups.

Note that our framework should serve as a general guide rather than an exhaustive set of principles for fair hiring. Some extensively studied dimensions of responsible AI, such as model training under particular fairness constraints, are not covered in this work. \revtext{Moreover, although this work addresses fairness concerns across a general population of job seekers without restricting job categories, contextual adaptation is necessary for specific job roles and hiring practices. For instance, in some experience-focused blue-collar jobs, online platforms may primarily support sourcing by connecting job seekers and recruiters rather than screening that relies on interview-based evaluation. In such cases, assessing input fairness, i.e., the composition of the applicant pool, becomes central, while certain design elements for algorithmic screening, such as nuanced variable selection, are less applicable. Similarly, for roles that emphasize portfolio evaluation over CVs (e.g., designers), design elements like ``nudges for objective evaluations'' should be tailored to the portfolio review process.} Additionally, it is important to consider normative values and value conflicts (e.g., the tension between system fairness and usability) before applying the framework to specific hiring contexts, as detailed in Section \ref{Dis-ValueTension}.


\section{DISCUSSION}

This work presents a comprehensive taxonomy of job seekers' fairness concerns in hiring, based on discourse in an online job community. Building on these insights, we outline a conceptual framework based on \modtext{VSD} that incorporates both algorithm and interface considerations for fair online recruitment. Overall, this work contributes to the HCI, CSCW, and responsible computing community by taking a human-centered approach to examining recruitment fairness, bridging the gap between real-world fairness challenges in hiring and practical implications for recruitment system design. 

\subsection{Unpacking Critical Gaps in Algorithmic Fairness through Value Sensitive Design}

The taxonomy of job seekers' fairness concerns addresses a critical yet often overlooked question: \textit{When developing a fair recruitment system, which fairness objectives should we focus on?} It highlights a range of fairness issues that are important to job seekers but are less considered as primary optimization objectives.

\subsubsection{Situating Fairness Concerns under Legal and Technical Contexts}


We emphasize a broader set of \revtext{personal attributes} beyond gender and ethnicity, which are the focus of most fair hiring algorithms~\cite{fabris2023fairness}. These attributes can be interconnected and can manifest through implicit and multi-modal proxies, challenging some existing debiasing strategies that may oversimplify the model~\cite{raghavan2020mitigating}. \modtext{Notably, what job seekers perceive as ``fair'' may go beyond what the law defines as discrimination. For example, EU non-discrimination law offers protection against discrimination in hiring on the basis of six grounds: age, disability, gender, religion or belief, racial or ethnic origin, and sexual orientation~\cite{fabris2023fairness,xenidis2020eu}, while our findings in Section \ref{RQ1-1} and \ref{RQ1-3} highlight additional concerns such as location, appearance, and employment gaps that are not easily captured even under the doctrine of \textit{indirect discrimination}~\cite{tobler2008limits}. Even for well-defined protected attributes, significant challenges remain, such as how to handle proxy attributes and how to ensure accountability in complex recruitment ecosystems. Operational difficulties therefore persist in translating legal principles into technical compliance of fair recruitment systems~\cite{xenidis2020eu,fabris2023fairness}. This gap calls for future interdisciplinary research that connects legal standards, technical auditing methods, and organizational practices to better align fairness protections with job seekers' lived concerns.} We suggest that policymakers actively reflect on the public's fairness concerns based on diverse channels such as social media and online communities, which may provide rich insights into refining the anti-discrimination policies and regulations.



Our findings also suggest that individuals are concerned about differential treatment due to specific interactions with hiring systems and recruiters. This underscores the importance of interactions in fair decision-making, \modtext{aligning with the current HCI focus on applying design interventions for fair decision-making~\cite{leung2020race,yang2024fair}. Notably, job seekers expect transparency and accountability from recruitment ecosystems, as shown in Section \ref{RQ1-2}, which often extend beyond the scope of standard algorithmic audits to the intertwined decisions of recruiters and algorithms. Future work should therefore develop hybrid frameworks that strengthen socio-technical auditing and enable tracing how outcomes emerge from the interaction between human recruiters and algorithmic tools. Such consideration is crucial to enhance the effectiveness of fairness monitoring (e.g., post-market monitoring under the EU AI Act~\cite{EU_AI_Act_Article_72}).}



\subsubsection{Gaps in Algorithmic Fairness through VSD}

Through VSD, the proposed conceptual framework provides an ideation scaffold that synthesizes job seekers' fairness concerns across algorithmic and interface levels. It responds to two key gaps in algorithmic fairness: (1) the use of a single, idealized fairness measure often limits what the algorithm is able to optimize for, preventing generalization to more diverse groups and broader conceptions of fairness~\cite{sarkar2024s} (the gap between \textbf{\textit{fairness conceptualization}} and \textbf{\textit{fairness operationalization}}); (2) computational approaches often fail to align mathematical fairness with job seekers' real-world fairness perceptions~\cite{lavanchy2023applicants} (the gap between \textbf{\textit{fairness perception}} and \textbf{\textit{fairness implementation}}). To this end, we use the taxonomy of fairness concerns as a lens to reflect on the entire algorithm development pipeline, including sample composition, feature selection, preprocessing, model selection, evaluation, deployment, and feedback loops.


\revtext{Beyond design implications for fair algorithms, we suggest policy makers and recruiters to critically examine the utility and necessity of adopting hiring algorithms, rather than defaulting to them as efficiency-enhancing tools - an assumption often associated with ``AI snake oil'' solutions~\cite{narayanan2025ai} with limited validity~\cite{rhea2022external}. As shown in Sec. \ref{RQ1-4}, algorithmic decision-making itself may embed the power imbalance between job seekers and platforms or recruiters. For example, applicants typically do not know which features are processed by the algorithm or how they are evaluated. Even design elements for algorithmic fairness (e.g., D3.2 \textit{establishing objective target variables} and D3.3 \textit{nuanced variable selection}) are difficult for users to operate or observe. Enhancing transparency and explainability may help mitigate this power imbalance; however, it may also amplify the risk of ``gaming the algorithm''~\cite{epstein2020will} and may introduce further inequities due to differences in applicants' algorithmic literacy. Although this work primarily aims to promote fair hiring in existing online recruitment systems, questions such as whether to adopt them, how to do so, and at what stage to deploy them remain central.
}


From a complementary perspective, the framework for designing fair recruitment interfaces \revtext{considers fair online recruitment beyond algorithmic decision-making, focusing on the role of interface-level mediation between humans and recruitment systems.} It aims to enhance hiring fairness in two ways: (1) for recruiters, it seeks to reduce factors contributing to conscious and unconscious human bias in hiring decisions~\cite{yang2024fair}; (2) for job seekers, it aims to minimize latent reasons for perceived unfairness by developing transparent and fair decision-making procedures~\cite{lee2019procedural,wang2020factors,shulner2023enhancing}. This framework provides practical design implications for all stages of hiring that involve interactions among recruiters, job seekers, and hiring systems, including sourcing, screening, selection, and evaluation, pointing toward future work to further evaluate and operationalize fair hiring interventions.

\subsection{Reflecting on the Value Tensions in Fair Recruitment Design}\label{Dis-ValueTension}

\subsubsection{Conflicts between Different Values}

\modtext{
Value conflicts are pervasive in value sensitive design (VSD)~\cite{miller2007value,felber2025addressing,zhu2018value,iqbal2021search}. In the algorithmic fairness literature, a well-known example is the conflict between individual fairness and group fairness, which embody different worldviews and are often mathematically incompatible~\cite{friedler2021possibility}. Correspondingly, users may hold divergent perceptions of how algorithmic decisions should balance competing values (e.g., equal treatment vs. equal impact~\cite{mourali2025public}). Our findings echo these contradictions: Some concerns emphasize \textit{group-level equity}, whereas others focus on the need for \textit{individualized consideration}; Some concerns necessitate \textit{standardized pipelines} to prevent unfair judgments arising from subjective impressions, whereas others emphasize the importance of \textit{contextual flexibility}. Given that algorithmic fairness often prioritizes a single optimization objective~\cite{fabris2023fairness}, we advocate for future work to adopt an iterative design that allows reaching a thoughtful compromise over competing values. For instance, while completely eliminating proxy features may reduce bias, it can also undermine predictive performance. A more balanced approach could involve feature audits and reweighting to constrain high-risk proxies while preserving job-relevant signals.}



\subsubsection{Value Conflicts between Stakeholders}

It is also important to pay attention to the values and incentives that motivate different parties in the recruitment market~\cite{lashkari2023finding}. Companies and job candidates may have conflicting success metrics driven by differing goals and incentives, creating situations that can appear unfair due to differing philosophies among parties, which leads to a systemic gap in fairness perceptions. \modtext{For example, our findings suggest job seekers demand \textit{transparency} in hiring decisions (e.g., whether and why they are rejected) as shown in Section \ref{RQ1-4}, but recruiting companies may intentionally or unintentionally keep \textit{opacity} to keep evaluation criteria flexible and potentially avoid legal liability; Job seekers expect \textit{data minimization} to reduce sensitive data or proxy attributes in Section \ref{RQ1-1}, yet recruiting companies sometimes collect unnecessarily rich data (such as personality evaluations) for \textit{comprehensive evaluation}; Some job seekers wish \textit{individualized consideration} with context-aware processes (such as considering their justifications for \revtext{negative background history}) in Section \ref{RQ1-3}, which may be hard to achieve when companies value \textit{efficiency} given large applicant pools. Beyond the recruiter–candidate relationship, value conflicts also arise between job seekers and hiring platforms that mediate recommendations, as shown in Section \ref{RQ1-4}. While job seekers expect \textit{relevance} and \textit{equitable exposure} in job recommendations, platforms often prioritize \textit{revenue models} and \textit{advertiser incentives}. In these cross-stakeholder conflicts, job seekers usually hold the weaker position, with limited influence over technical design. We therefore call for the development of regulatory frameworks that systematically identify and address such value conflicts to mitigate systemic misalignments. In addition, the fair hiring framework in this work is grounded in fairness concerns from the perspective of job seekers as directly impacted stakeholders. However, we emphasize that it is crucial to integrate multi-stakeholder perspectives before implementing specific designs, thereby balancing competing values and ensuring both feasibility and sustainability.
}

\section{LIMITATIONS}\label{limitations}

This work has the following limitations. First, we identified job seekers' concerns based only on r/jobs, one of the largest online job communities on Reddit. With its large community size and the absence of job-related constraints, it addresses fairness concerns across various stages of job seeking and captures the considerations of job seekers in a natural setting. However, it also inherits the limitations of online community analysis, such as representing only those willing (and able) to share job-seeking-related experiences in the community, as well as failing to surface in-depth user insights. Therefore, future work based on other human-centered approaches, such as large-scale surveys and in-depth interviews, can supplement the understanding of job seekers' fairness concerns for fair hiring design.


Additionally, Reddit remains a Western-centered platform, with nearly half of its users from the US. As an online community, it may also fail to adequately represent low-income, blue-collar, or older job seekers. \revtext{Therefore, we suggest further research to more systematically examine fairness concerns among these underrepresented groups, and call for more context-specific fair hiring investigations across different professional roles.}



\modtext{Finally, our analysis aimed to develop a conceptual framework for fair recruitment by distilling recurring patterns of fairness concerns across diverse contexts, domains, and tools, rather than providing context-specific causal diagnoses. \revtext{Our VSD processes encompassed both universal structural fairness issues (e.g., opaque recruitment processes) as well as those influenced by social and cultural factors (e.g., age bias may vary by cultural contexts), highlighting the importance of ensuring value alignment before deployment in real-world settings within specific cultural contexts.} In our framework, we show the corresponding empirical evidence and normative value dimensions when presenting the design implications in Section \ref{Finding-Implication} to support the assessment of design generalizability.} 



\section{CONCLUSION}

This study takes a value sensitive design approach to fair online recruitment, aiming to bridge the gap between theoretical fairness conceptualization in hiring system development and real-world fairness challenges in hiring practices. To achieve this, we first adopt a grounded theory approach to develop a comprehensive taxonomy of job seekers' fairness concerns in hiring from r/jobs, one of the largest online job communities. This taxonomy reveals issues such as interaction bias, improper interpretations of qualifications, and power imbalances that affect job seekers' perceptions of fairness in the recruitment process, highlighting bias factors that may be overlooked in the development of fair recruitment systems. Building on the taxonomy, we conduct value sensitive design with empirical, conceptual, and technical investigations, outlining a conceptual framework for online recruitment systems that spans every stage of fair algorithm development and recruitment interface design. In the process, we suggest how practitioners can more generally translate fairness issues raised by job seekers into concrete system designs.

\begin{acks}
This work is supported by the FINDHR project, Horizon Europe grant agreement ID: 101070212. 
\end{acks}

\bibliographystyle{ACM-Reference-Format}
\bibliography{sample-base}

\end{document}